\documentclass[12pt]{article}
\pdfoutput=1 

\usepackage{cancel,slashed}
\usepackage{bbm}
\usepackage{amssymb}
\usepackage{amsfonts}
\usepackage{amsmath}
\usepackage{graphicx}
\usepackage{cite}
\usepackage{latexsym}
\usepackage{color}
\usepackage{xypic}
\usepackage{cancel,slashed}
\usepackage[linktocpage]{hyperref}
\usepackage{color}
\usepackage{transparent}
\usepackage{footmisc}
\usepackage{float}

\makeatletter
\def\@xfootnote[#1]{%
  \protected@xdef\@thefnmark{#1}%
  \@footnotemark\@footnotetext}
\makeatother

\newcommand{\bmat}{\left(\begin{array}}
\newcommand{\emat}{\end{array}\right)}

\def\p{\partial}
\def\a{\alpha}

\def\b{\beta}
\def\g{\gamma}

\def\th{\theta}

\def\-{\hphantom{-}}
\def\ov{\overline}
\def\s2{\frac{1}{\sqrt2}}

\def\oh{\frac{1}{2}}
\def\beq{\begin{equation}}
\def\eeq{\end{equation}}
\def\beqa{\begin{eqnarray}}
\def\eeqa{\end{eqnarray}}

\def\im{{\rm Im \,}}
\def\re{{\rm Re \,}}

\def\T{{\rm T}}

\def\Dsl{\,\raise.15ex\hbox{/}\mkern-13.5mu D} 

\def\CH {{\cal H}}

\def\CN {{\cal N}}
\def\CF {{\cal F}}

\def\CS {{\cal S}}

\def\CO {{\cal O}}

\def\re{\mbox{Re}}
\def\im{\mbox{Im}}

\def\be{\begin{equation}}
\def\ee{\end{equation}}
\newcommand{\al}[1]{\begin{align}#1\end{align}}
\def\bea{\begin{eqnarray}}
\def\eea{\end{eqnarray}}
\def\raw{\rightarrow}
\def\Raw{\Rightarrow}

\def\IC{\mathbb{C}}
\def\IN{\mathbb{N}}
\def\IZ{\mathbb{Z}}
\def\IR{\mathbb{R}}

\def\T{{\bf T}}

\def\oh{\frac{1}{2}}
\def\a{{\alpha}}
\def\b{{\beta}}

\def\eps{{\epsilon}}
\def\th{{\theta}}
\def\Lam{{\Lambda}}
\def\lam{{\lambda}}

\def\sig{{\sigma}}
\def\Sig{{\Sigma}}
\def\g{{\gamma}}

\def\p{{\partial}}

\def\sm2{{\mbox{\small 2}}}

\begin{document}
\pagestyle{plain}

\makeatletter
\@addtoreset{equation}{section}
\makeatother
\renewcommand{\theequation}{\thesection.\arabic{equation}}
\pagestyle{empty}
\rightline{ IFT-UAM/CSIC-17-027}
\rightline{MAD-TH-17-02}
\vspace{0.5cm}
\begin{center}
\Huge{{Flux Flattening\\ in Axion Monodromy Inflation}
\\[7.5mm]}
\large{Aitor Landete,$^1$ Fernando Marchesano,$^1$ Gary Shiu,$^{2,3}$\\ and Gianluca Zoccarato$^{2,3}$ \\[7mm]}
\small{
${}^1$ Instituto de F\'{\i}sica Te\'orica UAM-CSIC, Cantoblanco, 28049 Madrid, Spain \\[1mm] 
${}^2$Department of Physics, University of Wisconsin, Madison, WI 53706, USA \\[1mm] 
${}^3$ Institute for Advanced Study, The Hong Kong University
  of Science and Technology, \\ Hong Kong, China
  \\[5mm]} 
\small{\bf Abstract} \\[5mm]
\end{center}
\begin{center}
\begin{minipage}[h]{15.0cm} 

String theory models of axion monodromy inflation exhibit scalar potentials which are quadratic for small values of the inflaton field and evolve to a more complicated function for large field values. Oftentimes the large field behaviour is gentler than quadratic, lowering the tensor-to-scalar ratio. This effect, known as flattening, has been observed in the string theory context through the properties of the DBI+CS D-brane action. We revisit such flattening effects 
in type IIB flux compactifications with mobile D7-branes, with the inflaton identified with the D7-brane position. We observe that, with a generic choice of background fluxes, flattening effects are larger than previously observed, allowing to fit these models within current experimental bounds. In particular, we compute the cosmological observables in scenarios compatible with closed-string moduli stabilisation, finding tensor-to-scalar ratios as low as $r \sim 0.04$. These are models of single field inflation in which the inflaton is much lighter than the other scalars through a mild tuning of the compactification data.

\end{minipage}
\end{center}
\newpage
\setcounter{page}{1}
\pagestyle{plain}
\renewcommand{\thefootnote}{\arabic{footnote}}
\setcounter{footnote}{0}


\tableofcontents


\section{Introduction}
\label{s:intro}

In recent years, observation data together with theoretical considerations are beginning 
to constrain the amplitude of primordial gravitational waves generated by inflation.
Combining the B-mode results from BICEP2 and KECK array CMB polarisation experiment with the (more model-dependent) constraints from Planck analysis of CMB temperature plus BAO and other data, yields a combined limit on the tensor-to-scalar ratio 
$r_{0.05}<0.07$ at 95\% confidence \cite{Array:2015xqh}. Through the Lyth bound \cite{Lyth:1996im}, this implies an experimental upper bound (under the assumptions behind \cite{Lyth:1996im}) on the inflaton field range.

At the same time, theoretical considerations
seem to suggest that certain large field inflationary models are in tension with quantum gravity \cite{ArkaniHamed:2006dz,delaFuente:2014aca, Rudelius:2015xta, Montero:2015ofa, Brown:2015iha, Brown:2015lia, Hebecker:2015rya, Bachlechner:2015qja, Junghans:2015hba, Heidenreich:2015wga}. These considerations take various forms, one of which is the weak gravity conjecture (WGC) \cite{ArkaniHamed:2006dz}. When applied to axion inflation, the WGC gives a correlated bound on the axion field range and the instanton action that generates the axion potential. 
Axion monodromy inflation \cite{Silverstein:2008sg,McAllister:2008hb} and in particular F-term axion monodromy inflation \cite{Marchesano:2014mla} (see \cite{Marchesano:2014mla,Blumenhagen:2014gta,Hebecker:2014eua,Ibanez:2014kia,Franco:2014hsa,Blumenhagen:2014nba,Hayashi:2014aua,Hebecker:2014kva,Ibanez:2014swa,Garcia-Etxebarria:2014wla,Blumenhagen:2015kja,Escobar:2015fda,Escobar:2015ckf,Hebecker:2015tzo,Bizet:2016paj,Landete:2016cix} for realisations) remains an interesting exception, as the inflaton potential is not only generated by instantons.\footnote{Though suppressing membrane nucleation still imposes a somewhat milder bound \cite{Ibanez:2015fcv,Hebecker:2015zss,Brown:2016nqt}.} Indeed, a monodromy axion is mapped to a massive gauge field through T-duality \cite{Brown:2015iha} so the WGC does not immediately apply.

Taking hints from these experimental findings and theoretical considerations, it is worthwhile to explore if string theory models of axion monodromy inflation naturally come equipped with mechanisms to lower the power of the tensor mode  compared to standard chaotic inflation.\footnote{Note, however, that we are not implying that these mechanisms necessary make the tensor mode unobservable, as detectable tensors require only $r \gtrsim {\rm (a~few)} \times 10^{-3}$.}  It has been argued that backreaction of the inflaton potential energy on heavy scalar fields can flatten a quadratic inflationary potential at large inflaton values \cite{Dong:2010in}. This ``flattening" then provides an example of a mechanism to lower $r$. However, whether flattening occurs depends on how the inflaton couples to the heavy fields, and hence a diagnostic is possible only if the UV completion of inflation is known.
For example, in string theory constructions with D-branes \cite{Baumann:2014nda,Westphal:2014ana,Silverstein:2016ggb}, flattening can follow from the structure of the DBI+CS action \cite{Silverstein:2008sg,McAllister:2008hb,Berg:2009tg,Dong:2010in,Gur-Ari:2013sba,Palti:2014kza,Marchesano:2014mla,Ibanez:2014swa}. However, the degree of flattening that one finds in this context is to date rather limited, e.g., a quadratic potential gets flattened to a linear potential through the $\alpha'$ effects included in the DBI action.

In this paper, we would like to point out a new source of flattening that we dub as {\it flux flattening}. This source of flattening is only visible for sufficiently large field ranges and hence it is not captured in the supergravity limit. Therefore, it represents an additional source of flattening to the effects seen in the supergravity literature. Despite having been overlooked, flux flattening effects entail strong flattening power, being able to lower the tensor power of a quadratic potential to well within the current experimental bound from combining the data from PLANCK, BICEP2/Keck Array and BAO.

We analyse flux flattening in the context of type IIB/F-theory flux compactifications with mobile 7-branes \cite{Hebecker:2014eua,Ibanez:2014kia,Ibanez:2014swa}, where this effect is easily described. Indeed, it is well known that in the presence of three-form background fluxes D7-branes experience a potential as we displace their position moduli from the vacuum. At small field values, such potential only depends on certain flux components, namely those that induce a non-supersymmetric B-field on the D7-brane worldvolume \cite{Camara:2004jj,Gomis:2005wc}. However, at large field values all background fluxes will contribute to the D7-brane energy, as one can see through direct evaluation of the DBI+CS action. Moreover, the kinetic term of a given position modulus will also depend on all these fluxes, resulting in an inflaton dependent kinetic term that will flatten the potential. The latter effect was already observed in \cite{Ibanez:2014swa} for a particular choice of background fluxes allowed by an orbifold projection. There, the growth of the kinetic terms with large inflaton values matched that of the potential, resulting in flattening to a linear potential. As we will show, once all background fluxes are taken into account the growth of the D7-brane position kinetic term will always be larger than that of its potential, thus inducing larger flattening effects than those observed in \cite{Ibanez:2014swa}. The functional dependence of the scalar potential that arises in this more general case has moreover a richer structure and interesting phenomenological features. 

In order to extract predictions from this more general setup it is important to relate the parameters of the D7-brane potential with the compactification data, and in particular with those compactifications that are compatible with the known schemes for closed-string moduli stabilisation. Therefore, we consider the embedding of our system into type IIB/F-theory compactifications with background fluxes. We find that in certain cases the scalar potential displays flat directions, in which the D7-brane position $\Phi$ and the 4d dilaton $S$ vary simultaneously. Such flat directions can be easily understood in terms of the symmetries of the effective K\"ahler and superpotential of the compactification. Then, by choosing slightly different superpotential parameters, one may engineer a very light direction that would represent the trajectory of inflation. As we will discuss, this scheme favours models of single field inflation, in which only one real component of the D7-brane moduli space is below the Hubble scale and the others stay at a mass scale similar to the K\"ahler moduli of the compactification. We use this fact to obtain an effective theory for $\Phi$ and a K\"ahler modulus $T$, valid below the scale of complex structure moduli masses. As in \cite{Bielleman:2016olv}, we use this effective theory to study moduli stabilisation and compute the backreaction effects of $T$ and the heavy component of $\Phi$ on the inflaton potential. 

Given this large single-field inflation scenario compatible with moduli stabilisation one may input its typical scales into the DBI+CS action. Interestingly, we find that the a priori complicated structure for the potential simplifies, and we recover a flux flattened potential that only depends on a single parameter, constrained to take values in a particular range. This in turn results in cosmological observables that cover the range of values $n_s \simeq 0.96 - 0.97$ and $r \simeq 0.04 - 0.14$, as show in figure \ref{fighat}.

This paper is organised as follows. In section \ref{sec:flatflux} we compute the D7-brane position potential for flux backgrounds that induce supersymmetric and non-supersymmetric worldvolume fluxes, generalising previous analysis. We describe the asymptotic behaviour of this potential at large field values and input the parameters from moduli stabilisation to compute the cosmological observables of the model. In section \ref{sec:embedding} we embed the mobile D7-brane as an F-term axion monodromy model in the context of type IIB/F-theory flux compactifications. We describe the discrete and continuous shift symmetries that appear in simple examples. We use the latter to formulate an scenario of single-field inflation with a realistic mass spectrum and compatible with K\"ahler moduli stabilisation. We finally  draw our conclusions in section \ref{sec:conclu}. Finally, appendix \ref{ap:SL} describes an alternative procedure to obtain an effective field theory for the inflaton field, and appendix \ref{ap:other} explores flux flattening effects for more general single field potentials.

\section{7-branes and flux flattening}
\label{sec:flatflux}

Following \cite{Hebecker:2014eua,Ibanez:2014kia,Ibanez:2014swa}, one may consider scenarios of large field inflation in which the inflaton candidates are D7-brane position moduli lifted by the presence of background fluxes. The potential generated for such moduli can be easily computed by means of 4d supergravity for small inflaton vevs but, as shown in \cite{Ibanez:2014swa}, in the regime of interest for inflation this approximation fails and one should compute the potential directly from the D7-brane action. This large-field computation was carried out in \cite{Ibanez:2014swa} for the restricted set of ISD background fluxes that respect the orbifold symmetry of the Higgs-otic setup, and generalised in \cite{Bielleman:2016olv} to include IASD fluxes respecting the same symmetry. 

In the following we would like to generalise the computation in \cite{Ibanez:2014swa} to include the more generic set of ISD background fluxes that will appear in general compactifications with mobile 7-branes like in \cite{Hebecker:2014eua}, and to consider varying dilaton and warp factors.  As we will see, while the effect of these extra fluxes does not appear in the scalar potential for small  7-brane displacements (and it is therefore invisible in the supergravity approximation) it produces an important flattening in the scalar potential for sufficiently large values of the 7-brane position modulus. Finally, although the computations will be performed by reducing the D7-brane action, by simply applying $SL(2,\IZ)$ duality it is easy to see that our conclusions apply to any mobile 7-brane.

\subsection{The closed string background}

Let us consider a type IIB/F-theory flux compactification with a 10d Einstein frame metric of the form
\be
\label{AnsE}
d s^2_{10}=Z^{-1/2}(y)d x^\mu d x_\mu+Z^{1/2}(y)\hat g_{mn}(y)d y^md y^n
\ee
where $\hat g$ is an F-theory three-fold metric on the internal space, with K\"ahler form $\hat J$ and holomorphic $(3,0)$-form $\Omega_0 = g_s^{1/2} \hat \Omega$, and $Z$ is the warping. As in \cite{Giddings:2001yu}, on top of this background there is a set of 7-branes sourcing a holomorphic axio-dilaton $\tau = C_0 + i g_s^{-1}$, D3-branes sourcing $Z$ and the self-dual RR flux $F_5$, and an imaginary-self-dual (ISD) three-form flux background $G_3 = F_3 - \tau H_3$.

Let us now look at a neighbourhood of a D7-brane wrapping a four-cycle $\CS$, and introduce local coordinates $(z_1,z_2,z_3)$ such that the D7-brane is localised in the $z_3$-plane. In such a region we consider an ISD primitive three-form flux $G_3$ of the form 
\al{\label{G3s2}
G_3 =   S_{\bar{1}\bar{1}} \, d \bar z_1 \wedge d z_2 \wedge dz_3 + S_{\bar{2}\bar{2}} \,d z_1 \wedge d \bar z_2 \wedge dz_3 +S_{\bar{3}\bar{3}}\,d z_1 \wedge d z_2 \wedge d \bar z_3 + G_{\bar{1}\bar{2}\bar{3}}\, d \bar z_1 \wedge d\bar z_2 \wedge d\bar z_3
}
where $S_{\bar{k}\bar{k}}$ and $G_{\bar{1}\bar{2}\bar{3}}$ are approximated to be constant. The important effect of the presence of the $G_3$ flux on the dynamics of the D7-brane will be given by the pullback of the B-field on its worldvolume. In particular in the proximity of the D7-brane we can integrate the relation $dB_2 = - \text{Im} \, G_3/\text {Im} \tau$, obtaining
\al{\label{B2}B_2 = -\frac{g_s }{2i} \Big[S_{\bar{1}\bar{1}}\,  z_3\, d \bar z_1 \wedge d z_2 + S_{\bar{2}\bar{2}} \, z_3\, d z_1 \wedge d \bar z_2 + S_{\bar{3}\bar{3}}\, \bar z_3\, d z_1 \wedge dz_2 +G_{\bar{1}\bar{2}\bar{3}}\,
\bar z_3\, d \bar z_1 \wedge d\bar z_2 - \text{h.c.}\Big]\,.
}
We may now identify the normal coordinate to the D7-brane with the brane position modulus via $z_3 = \sigma \Phi$, with $\sig =2\pi \a' = l_s^2/2\pi$. This implies that the pullback of the B-field on the worldvolume of the D7-brane, and therefore $\CF = B_2 -\sig F$,  will depend on its location. Since supersymmetry is achieved when $\mathcal F^{(0,2)} = 0$ on the D7-brane, we see that the flux components $S_{\bar{3}\bar{3}}$ and $G_{\bar{1}\bar{2}\bar{3}}$  will naturally stabilise the brane position modulus at loci where this condition is met, which for vanishing magnetic fluxes on the worldvolume of the D7-brane is attained at $B^{(0,2)} = 0$ or equivalently $z_3 = 0$.

In addition to the form of the $G_3$ flux we will need the values of the RR fluxes and potentials that enter the D7-brane Chern-Simons action. In particular we will need the following set of relations
\al{\label{eq:RRdual}
d C_6 -  H_3 \wedge C_4 &= - g_s \star_{10} \,\text{Re}\, G_3\, = \, - Z^{-1} d \text{vol}_{\mathbb R^{1,3}} \wedge H_3,\\
d C_8 - H_3 \wedge C_6 &=  g_s^2 \star_{10} \,\text{Re}\, d\tau\, = - \oh d \left(g_s\, d{\rm vol}_{\IR^{1,3}} \wedge \hat J \wedge \hat J \right),}
that can be obtained from the equations of motion. Finally we have that
\be
\tilde F_5\, =\,  dC_4 - \frac{1}{2} C_2 \wedge H_3 +\frac{1}{2}B_2 \wedge F_3  \, =\,  (1+ \star_{10} ) d\chi_4\,,
\ee
where
\be
\chi_4 = \chi \, d \text{vol}_{\mathbb R^{1,3}}, \quad \quad d\chi \, =\, dZ^{-1}.
\ee
With this at hand we proceed to compute the scalar potential felt by a D7-brane.

\subsection{The DBI+CS computation}


The Dirac-Born-Infeld (DBI) and Chern-Simons (CS) actions which control the dynamics of a single D7-brane are
\al{S_{DBI} &= - \mu_7  \int d^8 \xi\ g_s^{-1} \sqrt{- \text{det} (P[E_{MN} + \sigma F_{MN})}\,,\\
S_{CS} & = \mu_7 \int P \left[\sum_n C_{2n} \wedge e^{-B_2}\right] \wedge e^{\sig F}\,,
}
where $P[\cdot]$ denotes the pull-back on the worldvolume of the D7-brane and
\al{E_{MN} = g_s^{1/2} G_{MN} - B_{MN}\,, \quad \quad & \mu_7 = (2\pi)^{-3} \sigma^{-4}\, ,}
where $G$ is the 10d Einstein frame metric.

\paragraph{Dimensional reduction of the CS action.}

In order to evaluate the CS action of the D7-brane let us first consider how this action changes between two different D7-brane locations. That is, we consider a reference four-cycle $\CS_0$ and take a homotopic deformation $\CS$. Since both four-cycles lie in the same homology class there is a five-chain $\Sigma_5$ such that $\p\Sigma_5 = \CS - \CS_0$, and we have that
\bea 
\Delta S_{CS} & = & \mu_7 \int_{\IR^{1,3} \times \Sig_5} P \left[d\left(\sum_n C_{2n} \wedge e^{-B_2}\right) \right] \\ \nonumber
&= & \mu_7 \int_{\IR^{1,3} \times \Sig_5} \left(d C_8 - H_3 \wedge C_6 \right) - B_2 \wedge \left(d C_6 - H_3 \wedge C_4 \right) + \oh \tilde F_5 \wedge B_2 \wedge B_2 + \dots\\ \nonumber
& = &  \frac{\mu_7}{2} \int_{\IR^{1,3}} d{\rm vol}_{\IR^{1,3}} \, \int_{\Sig_5} d \left(Z^{-1}  B_2 \wedge B_2 - g_s  \hat J \wedge \hat J\right) 
\eea
where for simplicity we have turned off the gauge worldvolume flux $F$, and in the second line we have neglected terms that do not contribute to the chain integral. If in addition we assume that at $\CS_0$ the pull-back of $B_2$ vanishes and the volume contribution cancels with that of the remaining 7-branes we obtain that
\be
S_{CS}\,=\, \oh  \mu_7 \int_{\IR^{1,3}} d{\rm vol}_{\IR^{1,3}} \, \int_{\CS} \left( Z^{-1}  B_2 \wedge B_2 - g_s \hat J \wedge \hat J \right)
\ee

\paragraph{Dimensional reduction of the DBI action.}

To dimensionally reduce the DBI action we may follow a procedure similar to the one outlined in \cite{Ibanez:2014swa}. We arrive at the result
\al{\label{eq:DBIred}S_{DBI} = - \mu_7   \int_{\IR^{1,3} \times \CS}  \hspace*{-.75cm} d^8 \xi \, g_s \sqrt{\text{det}(g_{ab}) f(\mathcal F) \left[1+2 Z \sigma^2 \p_\mu \Phi \p^\mu \overline \Phi+\frac{1}{2}g_s^{-1} Z \sigma^2 F_{\mu \nu} F^{\mu \nu} \right]}\,,}
where by $\Phi $ we denote the complexified brane position modulus. The function $f(\mathcal F)$ appearing in \eqref{eq:DBIred} is defined as
\al{f(\mathcal F) = 1+ \epsilon \, \mathcal F^2 + \frac{1}{4} \epsilon^2 (\mathcal F \wedge \mathcal F)^2 \,,}
where $\epsilon = Z^{-1} g_s^{-1}$ and the contractions are made with the unwarped metric $\hat{g}_{ab}$ of $\CS$. Note that, since we are considering more general fluxes than the case appearing in \cite{Ibanez:2014swa}, the function $f(\mathcal F)$ is not a perfect square. Retaining only terms quadratic in derivatives we obtain the following terms from the DBI action
\al{S_{DBI} =  \mu_7   \int_{\IR^{1,3}} d{\rm vol}_{\IR^{1,3}} \, \int_{\CS}  \frac{g_s}{2} \hat J \wedge \hat J \, \sqrt{f(\mathcal F)} \left[  1 + Z \sigma^2 \p_\mu \Phi \p^\mu \overline \Phi+\dots\right]\,,
}
where we have used that the pull-back of $-\oh \hat J \wedge \hat J$ is the volume form of a holomorphic four-cycle like $\CS$, and where  the dots include higher derivative terms as well as terms involving the gauge field on the D7-brane.

\paragraph{The brane position modulus effective action.}

Let us summarise the 4d effective action controlling the dynamics of the brane position modulus. Adding up the DBI and CS contribution we obtain
\al{S_{\Phi} = -\int_{\mathbb R^{1,3}} d{\rm vol}_{\IR^{1,3}} \Big[  g(\CF)\p_\mu \Phi \p^\mu \overline \Phi + V(\CF)\Big]\,,}
where
\al{g(\CF) &= \frac{1}{(2\pi)^3\sig^2} \int_{\CS} g_s Z\, \sqrt{f(\mathcal F)}\, d\hat{\rm vol}_{\CS} \, , \\
V(\CF) &= \mu_7 \int_{\CS} g_s  \left[\sqrt{f(\mathcal F)} - 1 \right]\, d\hat{\rm vol}_{\CS} - \oh Z^{-1} \CF \wedge \CF \, ,}
and $d\hat{\rm vol}_{\CS}$ is the unwarped volume form of the D7-brane four-cycle. We may now perform the 4d Weyl rescaling 
\be
g_{\mu\nu} \, \raw\, \frac{g_{\mu\nu}}{{\rm Vol}_{X_6}}
\ee
with ${\rm Vol}_{X_6}$ is the volume of the compactification manifold $X_6$ in units of $l_s = 2\pi \sqrt{\a'}$. After that, mass scales in Planck units should be measured in terms of $\kappa_4^{-1} = \sqrt{4\pi} l_s^{-1}$ and the above quantities read
\al{g(\CF) &= \frac{1}{2\pi {\rm Vol}_{X_6}} \frac{1}{l_s^4} \int_{\CS} g_s Z\, \sqrt{f(\mathcal F)}\, d\hat{\rm vol}_{\CS} \, , \\
\kappa_4^4\, V(\CF) &= \frac{1}{8\pi {\rm Vol}_{X_6}^2} \frac{1}{l_s^4} \int_{\CS} g_s  \left[\sqrt{f(\mathcal F)} - 1 \right]\, d\hat{\rm vol}_{\CS} - \oh Z^{-1} \CF \wedge \CF \, ,}

Notice that if $\CF$ is a self-dual or anti-self-dual two-form in $\CS$ then
\be
\CF \wedge \CF = \pm \CF^2 d\hat{\rm vol}_{\CS} \quad  \Raw \quad f(\CF)\, =\, \left(1 + \oh \eps \CF^2 \right)^2 
\label{square}
\ee
and so in the former case the potential vanishes while in the latter we have
\be
 \kappa_4^4\, V(\CF) = \frac{1}{8\pi {\rm Vol}_{X_6}^2} \frac{1}{l_s^4} \int_{\CS} Z^{-1} \CF^2 \, d\hat{\rm vol}_{\CS}
\label{potasd}
\ee
as obtained in \cite{Gomis:2005wc}. The kinetic term and potential depend on $\Phi$ through eq.(\ref{B2}) and the identification $z_3 = \sigma \Phi$. To make this dependence more explicit let us turn off the worldvolume flux $F$ and introduce a new normalisation for the brane position modulus
\al{\label{eq:tildephi} \Phi \ \raw \  \left(\frac{\tilde{\mathcal V}_{\CS_0}}{2\pi {\rm Vol}_{X_6}}\right)^{-1/2} \Phi}
where 
\be
\tilde{\mathcal V}_{\CS}\, =\, \frac{1}{l_s^4} \int_{\CS} g_s Z\, d\hat{\rm vol}_{\CS}
\ee
and $\CS_0$ is the reference four-cycle where $P[B_2]$ vanishes, hence the minimum of the potential that corresponds to $\Phi=0$.  Note that with this choice of normalisation $\Phi $ has canonical kinetic terms at this minimum. After this redefinition we find the kinetic term and potential take the form 
\al{g(\Phi) &=  \frac{1}{ \tilde{\mathcal V}_{\CS_0}} \frac{1}{l_s^4} \int_{\CS} g_s Z \left[1+ \hat \eps\,(\mathcal G + \mathcal H)+\frac{1}{4} \hat \eps^2\,(\mathcal G - \mathcal H)^2  \right]^{\frac{1}{2}} \, d\hat{\rm vol}_{\CS} \,, \label{gPhi}\\
\kappa_4^4\, V(\Phi ) &=   \frac{1}{8\pi {\rm Vol}_{X_6}^2} \frac{1}{l_s^4} \int_{\CS} g_s \left(\left[1+ \hat \eps\,(\mathcal G + \mathcal H)+\frac{1}{4}\hat \eps^2\,(\mathcal G - \mathcal H)^2  \right]^{\frac{1}{2}}+ \frac{1}{2}
\hat\eps\,\mathcal G-\frac{1}{2}\hat \eps\, \mathcal H-1\right)\, d\hat{\rm vol}_{\CS} \,, \label{VPhi}
}
where we have defined
\be
\hat \eps = g_s \frac{2\pi  {\rm Vol}_{X_6}}{4Z \tilde{\mathcal V}_{\CS_0}}
\ee
and $\CH$ and ${\cal G}$ stand for the self-dual and anti-self-dual components of $P[B_2]$, respectively. Given (\ref{B2}) they read
\be
\mathcal G = |\overline G_{\bar 1\bar 2\bar 3}  \Phi - S_{\bar 3 \bar 3} \overline \Phi|^2\,, \quad \mathcal H = |S_{\bar 2 \bar 2} \Phi -\overline S_{\bar 1 \bar 1} \overline \Phi|^2\,.
\ee
In order to compare with the results in \cite{Ibanez:2014swa} let us consider that $g_s$ and $Z$ are constant.\footnote{Despite this simplification it could still happen that $g_s$ does depend on $\Phi$, which would complicate the functional dependence of $g(\Phi)$ and $V(\Phi)$. The effect of flux flattening discussed below would nevertheless still remain.} Then $\tilde{\mathcal V}_{\CS} = \tilde{\mathcal V}_{\CS_0}$ for any $\CS$ and so these expressions reduce to 
\al{g(\Phi) &=    \left[1+ \hat \eps\,(\mathcal G + \mathcal H)+\frac{1}{4} \hat \eps^2\,(\mathcal G - \mathcal H)^2  \right]^{\frac{1}{2}}\,,\\
\kappa_4^4\, V(\Phi ) &= \frac{\tilde{\mathcal V}_{\CS}}{8\pi  {\rm Vol}_{X_6}^2  Z}   \left(\left[1+ \hat \eps\,(\mathcal G + \mathcal H)+\frac{1}{4}\hat \eps^2\,(\mathcal G - \mathcal H)^2  \right]^{\frac{1}{2}}+ \frac{1}{2}
\hat\eps\,\mathcal G-
\frac{1}{2}\hat \eps\, \mathcal H-1\right)\,. \label{eq:scpot}
}
Note that if we set ${\mathcal H} = 0$ we recover the results in \cite{Ibanez:2014swa}. On the contrary, if ${\mathcal H} \neq 0$ we have that $[g(\Phi)]^2$ no longer is a perfect square and that $g$ and $V$ depend on quite different functions of $\Phi$.

Finally, in order to analyse the potential it is convenient to move to a different parametrisation for the brane position modulus. Specifically we may switch to polar coordinates in the plane normal to the D7-brane location and define 
\begin{subequations}
\begin{align}
\rho^2  &  = \Phi \overline \Phi\, \kappa_4^{-2}\\
A  & = 2|G_{\bar 1 \bar 2 \bar 3}S_{\bar 3 \bar 3}|/(|G_{\bar 1 \bar 2 \bar 3}|^2 + |S_{\bar 3 \bar 3}|^2) \\
\tilde A  & = 2|S_{\bar 1 \bar 1} S_{\bar 2 \bar 2}|/(|S_{\bar 1\bar 1}|^2 + |S_{\bar 2 \bar 2}|^2) \\ 
\theta  & = 2 \text{Arg}\, \Phi - \text{Arg}\, G_{\bar 1 \bar 2 \bar 3}S_{\bar 3 \bar 3} \\
\zeta &  =  \text{Arg}\, G_{\bar 1 \bar 2 \bar 3}S_{\bar 3 \bar 3} - \text{Arg}\, S_{\bar 1 \bar 1} S_{\bar 2 \bar 2}. 
\end{align}
\end{subequations}
The quantities $\mathcal G $ and $\mathcal H$ then simplify with this notation and become
\al{\mathcal G = \kappa_4^2 (|G_{\bar 1 \bar 2 \bar 3}|^2 + |S_{\bar 3 \bar 3}|^2) \Big[1- A \cos \theta \Big]\,\rho^2 \,, \quad \mathcal H = \kappa_4^2(|S_{\bar 1 \bar 1}|^2 + |S_{\bar 2 \bar 2}|^2) \Big[1- \tilde A \cos (\theta+ \zeta) \Big]\,\rho^2 \,.}

\subsection{Potential asymptotics and flux flattening}
\label{ssec:asympto}

\begin{figure}[H]
\begin{center}
\includegraphics[width=100mm]{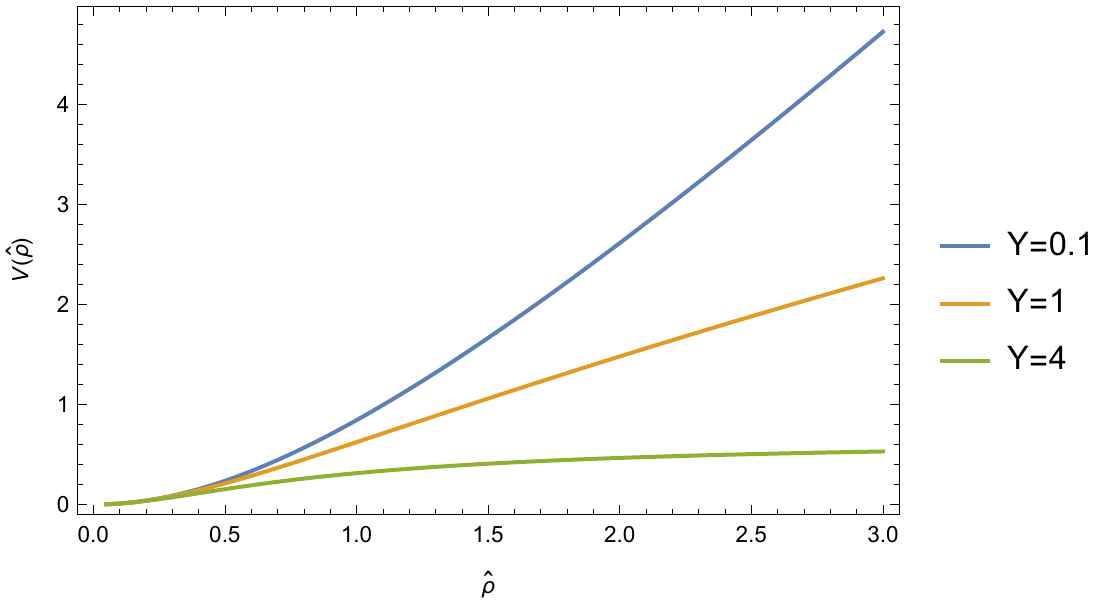}
\caption{The single field scalar potential for the canonically normalised inflaton $\hat \rho$ for different values of $\Upsilon$ keeping fixed $\hat G = 1$.}\label{fig1}
\end{center}
\end{figure}

Let us now turn to the analysis of the asymptotic behaviour of the above scalar potential. In order to compare with the large-field linear behaviour found in  \cite{Ibanez:2014swa} we again consider the simplified version \eqref{eq:scpot}, and for convenience we define the following quantities
\al{ \tilde  G =\hat \epsilon\, (|G_{\bar 1 \bar 2 \bar 3}|^2 +|S_{\bar 3 \bar 3}|^2) \kappa_4^2\,, \quad \Upsilon=\frac{|S_{\bar 1 \bar 1}|^2 +|S_{\bar 2 \bar 2}|^2 }{|G_{\bar 1 \bar 2 \bar 3}|^2 +|S_{\bar 3 \bar 3}|^2}\,.
}
The important parameter in the upcoming analysis will be $\Upsilon$, which measures the strength of supersymmetric components of the B-field induced on the D7-brane vs the non supersymmetric ones, and it will parametrically control the flattening of the scalar potential.
To gain an intuition over the asymptotics of the scalar potential we will consider regions in the parameter space where we effectively achieve single field inflation, as one of the components of $\Phi$ is much heavier than the other one. As we will see in section \ref{sec:embedding} and also pointed out in \cite{Bielleman:2016olv}, this limit seems favoured when embedding our D7-brane system in a setup with full moduli stabilisation. These cases admit an unified description and the shape of the potential will depend on two 
parameters, one the aforementioned $\Upsilon$ and the other which we choose to call $\hat G$ to be defined for each case. The cases we look into are the following 
two:
\begin{itemize}
\item[-] \textbf{Single field I.} Here we take $A = \tilde A=0$ so that the angular variable $\theta$ disappears from the potential. The inflaton is identified with the 
radial variable $\rho = \sqrt{\Phi \overline \Phi} \kappa_4^{-1}$ and in this case $\hat G = \tilde G$.

\item[-] \textbf{Single field II.} Here we take $A = \tilde A \simeq 1$ and $\zeta = 0$. Now the inflaton is the real part of $\Phi' = e^{-i \gamma/2} \Phi$ where 
$\gamma = \text{Arg} (G_{\bar 1 \bar 2 \bar 3} S_{\bar 3 \bar 3})$. Due to the fact that $A$ is very close to 1 the imaginary part of $\Phi'$ will have a much higher mass as compared to the real part. Therefore considering trajectories where the inflaton is $\re\, \Phi'$ is the inflaton and $\im\, \Phi'$ is frozen at the origin is a good approximation and the model becomes a single field model to all effects. In this case $\hat G = (1-A) \tilde G$.\footnote{In the limiting case where $A = \tilde A = 1$ and $\zeta = 0$, $\re\, \Phi'$ becomes a flat direction and one could see $\im\, \Phi'$ as driving single field inflaton, as considered in \cite{Ibanez:2014swa}. In that case one should take $\hat G = \tilde G$.} 

\end{itemize}
Both cases have in common that along the trajectories described it occurs that $\mathcal G = \hat G \rho^2 $ and $ \mathcal H = \Upsilon \hat G \rho^2$,
where $\rho$ stands for the inflaton field. Therefore the potential is identical in both and we can discuss its asymptotic shape at the same time.
The scalar potential we obtain is
\al{\frac{V(\rho)}{V_0} = \sqrt{1+\hat{G} (\Upsilon +1)\rho 
   ^2+\frac{1}{4} \hat{G}^2 (\Upsilon -1)^2  \rho ^4 }+\frac{1}{2}\hat{G} (1-\Upsilon)\rho ^2 -1\,, \label{fracV}}
where $\kappa_4^4 V_0 = \tilde{\mathcal V}_{\CS}(8\pi  {\rm Vol}_{X_6}^2 Z)^{-1}  $. We can easily analyse the asymptotic behaviour of the scalar potential for $\rho \rightarrow \infty$. The result turns
out to heavily depend on the value of $\Upsilon$
\al{\lim_{\rho \rightarrow \infty}\frac{V(\rho)}{V_0}=\left\{\begin{array}{l l}
\hat G(1-\Upsilon) \rho^2\,, & 0\leqslant\Upsilon <1\,,\\[2mm]
\sqrt{2\hat G}\, \rho & \Upsilon = 1\,,\\[2mm]
\dfrac{2}{\Upsilon-1}-\dfrac{4 \Upsilon}{\hat G \rho^2 (\Upsilon-1)^3}\,,  & \Upsilon >1\,.\\\end{array}\right.
}
We see therefore that if $\Upsilon>1$ -- namely when the strength of the self-dual B-field components is larger than the anti-self-dual ones -- the potential will approach a constant value as $\rho$ draws nearer to infinity. The resulting potential in this regime exhibits a plateau-like shape
and inflationary models constructed using this scalar potential will have a much lower value of tensor-to-scalar ratio as opposed to the usual power-law like potentials. So far
we have discussed the effect of flattening in the scalar potential, however as already noted in \cite{Ibanez:2014swa} additional flattening in the scalar potential will appear when 
considering the effect of the non trivial kinetic terms. To obtain the canonically normalised inflaton field $\hat \rho$ it is necessary to solve the integral equation
\al{\hat \rho = \int^{\rho} g^{1/2}(\rho') d\rho'\,,
\label{canonicalv}}
and invert the relation between $\hat \rho$ and $\rho$. Given the complexity of the kinetic terms we find it possible to attain canonical normalisation only numerically. Nevertheless
we can gain some intuition looking at large values of the inflaton field where the kinetic terms drastically simplify
\al{\lim_{\rho\rightarrow \infty}K_{\rho\rho} = \left\{\begin{array}{ll}\dfrac{1}{2} \hat G\, |\Upsilon-1|\,\rho^2 & \Upsilon \neq 1\,,\\[2mm] 
\sqrt{2 \hat G} \,\rho& \Upsilon = 1\,,\end{array}\right.
}
which yields the following potential for large values of the inflaton field in terms of the canonically normalised field
\al{\lim_{\hat \rho \rightarrow \infty}\frac{V(\hat \rho)}{V_0}=\left\{\begin{array}{l l}
 \sqrt{8\hat G}\,\dfrac{(1-\Upsilon)}{\sqrt{|\Upsilon-1|}} \,\hat \rho\,, & 0\leqslant\Upsilon <1\,,\\[2mm]
\left( \dfrac{9 }{2}\right)^{\frac{1}{3}}\, \hat G^{\frac{1}{3}}\, \hat {\rho}^{\frac{2}{3}} & \Upsilon = 1\,,\\[2mm]
\dfrac{2}{\Upsilon-1}-\dfrac{\sqrt{2 |\Upsilon-1|} \,\Upsilon}{\sqrt{\hat G } \,\hat \rho\, (\Upsilon-1)^3}\,,  & \Upsilon >1\,.\\\end{array}\right.
}
We chose to 
plot the form of the scalar potential for the canonically normalised inflaton field $\hat \rho$ for different values of $\Upsilon$ in figure \ref{fig1} to show more explicitly the flattening 
effect in the scalar potential.

Let us stress that this strong flattening effect will be absent in the supergravity discussion that we will carry in the next section, which will able to capture the inflaton scalar potential only in the regime of small values for $\rho$. Nevertheless, such a supergravity analysis will allow us to draw up an estimate for the typical values of the parameter in the DBI potential, as we discuss in the following.

\subsection{Estimating the scales of the model}
\label{ssec:estimate}

Let us briefly discuss a DBI potential compatible with the compactification scheme discussed in section \ref{sec:embedding}, and which considers the interplay of the D7-brane position modulus with the closed string moduli of the compactification. In particular, in subsection \ref{ssec:moduli} we will argue that a simple way to reproduce a scalar mass spectrum compatible with large field inflation and moduli stabilisation is by having one of the two components of the complex field $\Phi$ much lighter than the other one. Therefore, we will recover a single field inflation model with a  potential of the kind discussed above, and the details from the compactification will translate into some specific values for the parameters $V_0$, $\hat{G}$ and $\Upsilon$. In the following we would like to consider those typical values for $V_0$, $\hat{G}$ and $\Upsilon$ that are compatible with a realistic scalar mass spectrum and the moduli stabilisation scheme discussed in section \ref{ssec:moduli}, in order to obtain a constrained range of cosmological observables in the next subsection. 

First, we have that for small values of $\rho$ the potential becomes
\be
V(\rho) \, =\, V_0\, \hat{G} \rho^2 + \dots
\ee
with
\be
\kappa_4^4 V_0 \sim \frac{g_s {\rm Vol}_{\CS}}{8\pi  {\rm Vol}_{X_6}^2} \sim 4 \times \left( 10^{-6} - 10^{-5} \right)\, ,
\ee
where we have taken $g_s {\rm Vol}_{\CS} \sim 1-10$ and ${\rm Vol}_{X_6}^2 \sim 10^4$, the latter being a typical value compatible with the hierarchy of mass scales discussed in subsection \ref{ssec:moduli}, see e.g. footnote \ref{scales}. Comparing with the estimated mass for the inflaton near the vacuum we have that
\be
\kappa_4^4 V(\rho) \, \simeq \,  4 \times 10^{-11} \rho^2 \quad \Raw \quad \hat{G} \, \sim\, 10^{-6} - 10^{-5} \, .
\ee

Moreover, we have that $\Upsilon$ is the quotient between two different kind of fluxes. On the one hand $G_{\bar 1 \bar 2 \bar 3}$ and $S_{\bar 3 \bar 3}$ are fluxes that enter the inflaton scalar potential even at small field. On the other hand, $S_{\bar 1 \bar 1}$ and $S_{\bar 2 \bar 2}$ will be fluxes to which the D7-brane will be insensitive near the vacuum. However, these fluxes will be sensed by the complex structure moduli, to which they will give masses. Hence, unless $\Upsilon$ is constrained by some specific feature of the compactification,\footnote{More precisely, $\Upsilon$ could be constrained to vanish by an orbifold symmetry like in \cite{Ibanez:2014swa} or by the fact that $h^{1,1}(\CS) = 1$, see the discussion in section \ref{ss:periodic}.} one may estimate $\Upsilon^{1/2}$ as the quotient between the typical complex structure moduli mass (that is, the flux scale) and the mass of a D7-brane modulus. If we now focus on the single field scenario considered in section \ref{ssec:moduli}, which corresponds to the single field case II discussed above, and look at the mass relations found in section \ref{ssec:moduli}, we have that $\Upsilon^{1/2}$ is roughly the quotient between the flux scale and the mass of the heaviest component of the D7-brane modulus, namely $\im \, \Phi'$. In other words we have that
\be
\Upsilon \, \sim \, \frac{m_{\rm flux}^2}{m_{{\rm Im}\Phi'}^2}\, \sim\, \frac{N^2}{\kappa_4^2 |W_0|^2} \sim 10^2 - 10^{3} \, ,
\ee
where $N \in \IZ$ is the typical value of flux quanta, which we have taken around $N^2 \sim 1-10$. Finally, $W_0$ is as defined in subsection \ref{ssec:moduli}, from where we have taken the typical value $\kappa_4 W_0 \sim 0.1$.  

\begin{figure}[H]
	\begin{center}
\includegraphics[width=120mm]{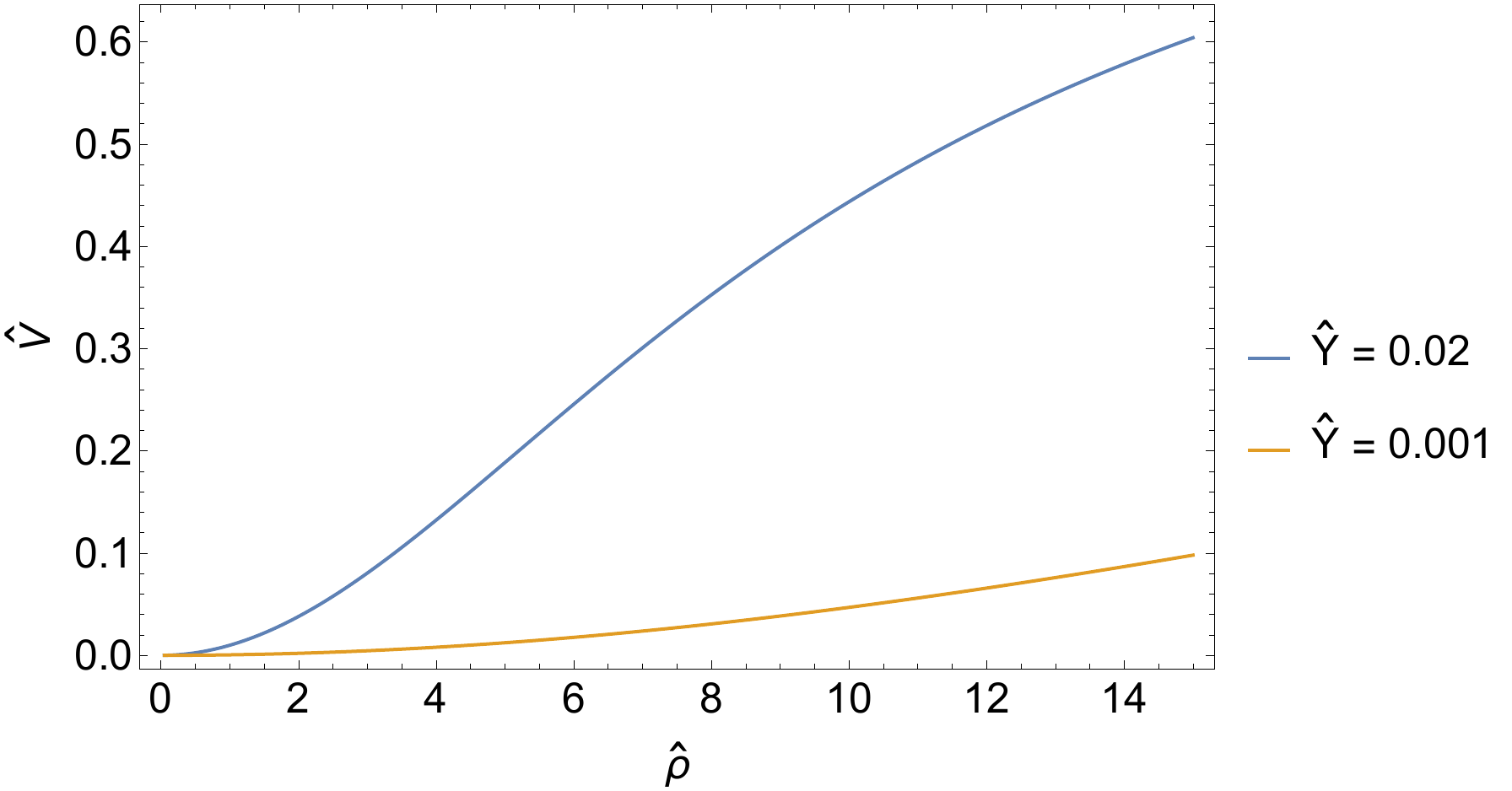}
\caption{Scalar potential $\hat{V}$ for the canonical field $\hat{\rho}$ for two different values of $\hat{\Upsilon}$.}\label{hatV}
\end{center}
\end{figure}

Given this large value of $\Upsilon$ and the small value of $\hat{G}$, we may approximate (\ref{fracV}) by
\be\label{eq:simplV}
\frac{V(\rho)}{V_0} = \frac{\hat{G} \rho^2}{1 + \oh\hat{G}  (\Upsilon -1) \rho^2} + \dots
\ee
so asymptotically
\be
V(\rho)\quad \stackrel{\rho \raw \infty}{\longrightarrow} \quad 2 V_0 \Upsilon^{-1} \, \sim\, (10^{-9} - 10^{-7}) \, \kappa_4^{-4}
\label{asympt}
\ee
which is intriguingly close to the scale of large-field inflation $V_{\text{inf},\star}^{1/4} = (10r)^{1/4} 1.88\times10^{16}\text{GeV}$ \cite{Ade:2015lrj}. This asymptotic constant value will not be changed by the field-dependent inflaton kinetic term, which for this choice of parameters can be approximated to be
\be\label{eq:simplK}
g(\rho)\, =\, 1 + \oh\hat{G}  (\Upsilon -1) \rho^2 + \dots\, .
\ee
Using (\ref{canonicalv}) we have that the canonically normalised field is given by
\be
\hat{\rho}\, =\,  \frac{\rho}{2}  \sqrt{1 + \oh \hat{G} (\Upsilon-1)  \rho^2}+\frac{ \sinh ^{-1}\left(\sqrt{\oh\hat{G}(\Upsilon-1)\rho^2}\right)}{\sqrt{2\hat{G}(\Upsilon-1)}} \, .
\ee
Hence, in the region where $\hat{G}  (\Upsilon -1) \rho^2 \ll 2$ we have that $\hat{\rho}  \simeq \rho$ and that (\ref{eq:simplV}) is a quadratic potential, and in the large field limit we have that $\hat{\rho} \simeq \sqrt{\frac{1}{8} \hat{G}  (\Upsilon -1)} \rho^2$ and that the potential asymptotes to the constant value (\ref{asympt}). In any event notice that for this range of parameters the potential can be written as
\be
V (\hat{\rho}) \, =\, \hat{V}_0 \cdot \hat{V} (\hat{\rho})
\label{hatpot}
\ee
where $\hat{V}_ 0 = 2 V_0 / (\Upsilon -1)$ and $\hat{V}$ is a monotonic function that only depends on the parameter $\hat{\Upsilon} = \hat{G}(\Upsilon-1)$, such that $\hat{V} \simeq \oh \hat{\Upsilon}  \rho^2$ at small field and asymptotes to $1$ for $\hat{\rho} \raw \infty$. In figure \ref{hatV} we plot $\hat{V}$ for some typical values of this parameter, within the range $\hat{\Upsilon} \sim 10^{-4}-10^{-2}$.
%

\subsection{Cosmological observables}
\label{ssec:cosmob}

Let us now analyse in some detail the cosmological observables that can be derived from the potential discussed above. In the single field scheme of subsection \ref{ssec:moduli} one finds that the distortion effect coming from the stabilisation of other moduli is sufficiently suppressed, and therefore the DBI+CS potential discussed in this section is a good approximation during the field ranges where inflation occurs.\footnote{More precisely, we find negligible backreaction effects from the heavy component of $\Phi$ and  K\"ahler moduli in the 4d supergravity model describing a mobile D7-brane, and we expect the same conclusion to apply to the flux-flattened DBI potential.}  Therefore, in the following we will focus on the single field scalar potential (\ref{hatpot}) and derive the phenomenological features of this model. We will see that even in this concrete case there is a rich phenomenology allowing for the possibility of having a moderately low tensor-to-scalar ratio. One may also analyse the features of single field D7-brane potential for other choices of parameters that may occur in different setups, as we do in appendix \ref{ap:other}.

As it usually happens for single field inflation to obtain the main cosmological observables, the spectral index $n_s$ and the tensor-to-scalar ratio $r$, it is sufficient to obtain the 
slow-roll parameters $\eta$ and $\epsilon$. For a single scalar field $\phi$ with non-canonical kinetic terms the slow-roll parameters are
\al{ \epsilon &= \frac{M_P^2}{2} \,G^{\phi \phi} \left(\frac{D_{\phi}V}{V}\right)^2\,,\\
\eta &= M_P^2\, G^{\phi \phi} \frac{D_{\phi}D_{\phi}V}{V}\,.
}
where $G^{\phi \phi}$ is the inverse of the target space metric and derivatives are covariant derivatives with the connection derived from the metric $G_{\phi \phi}$. Knowledge of
the slow-roll parameters is sufficient to compute cosmological observables: we copy here the well-known relations
\al{ n_s &= 1+2\eta_*-6\epsilon_*\,,\\
r &= 16 \epsilon_*\,,
}
where $\eta_*$ and $\epsilon_*$ are the values of $\eta$ and $\epsilon$ at the beginning of inflation. 

\begin{center}
\begin{figure}[H]
\includegraphics[width=80mm]{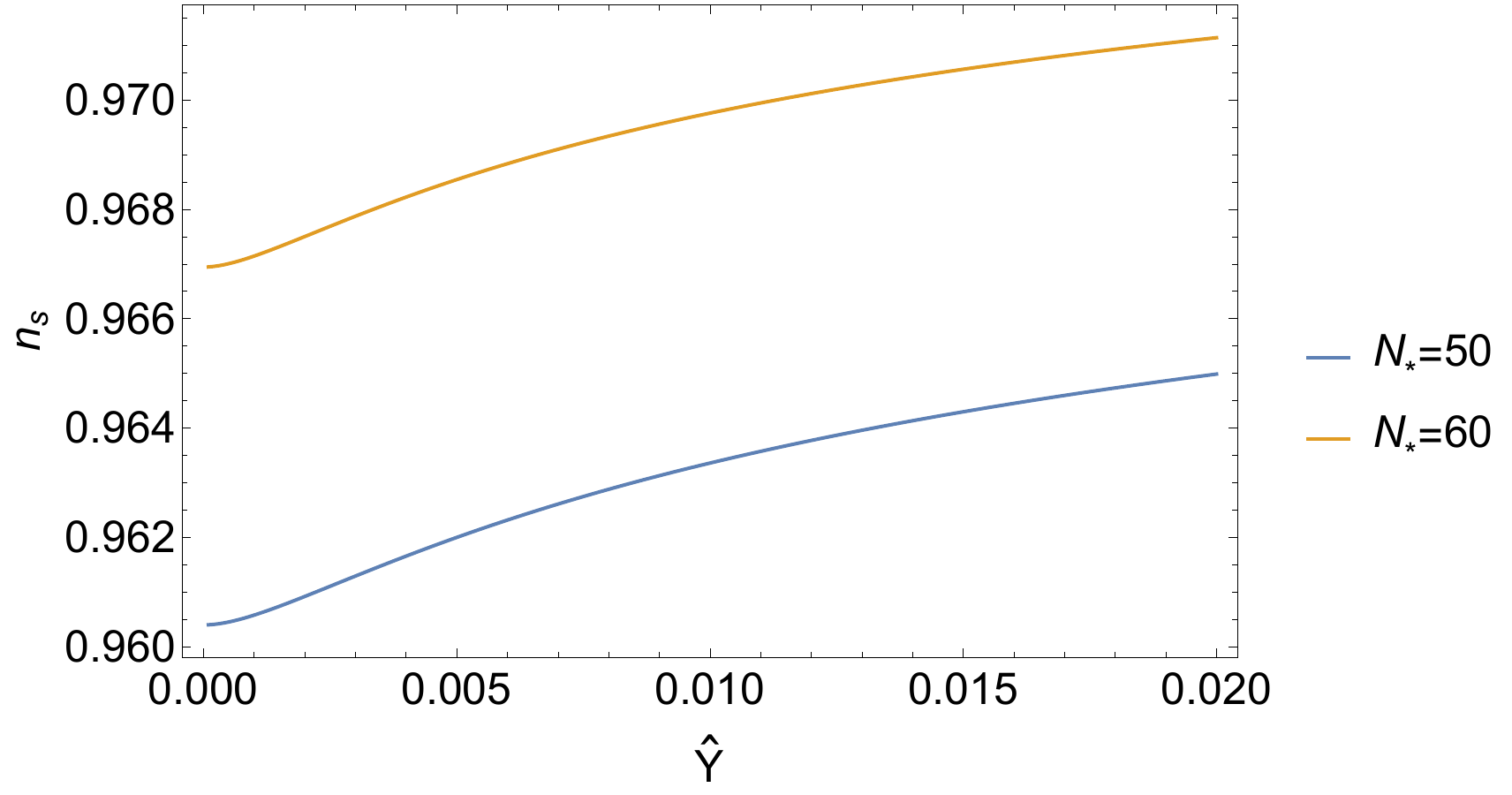} \includegraphics[width=80mm]{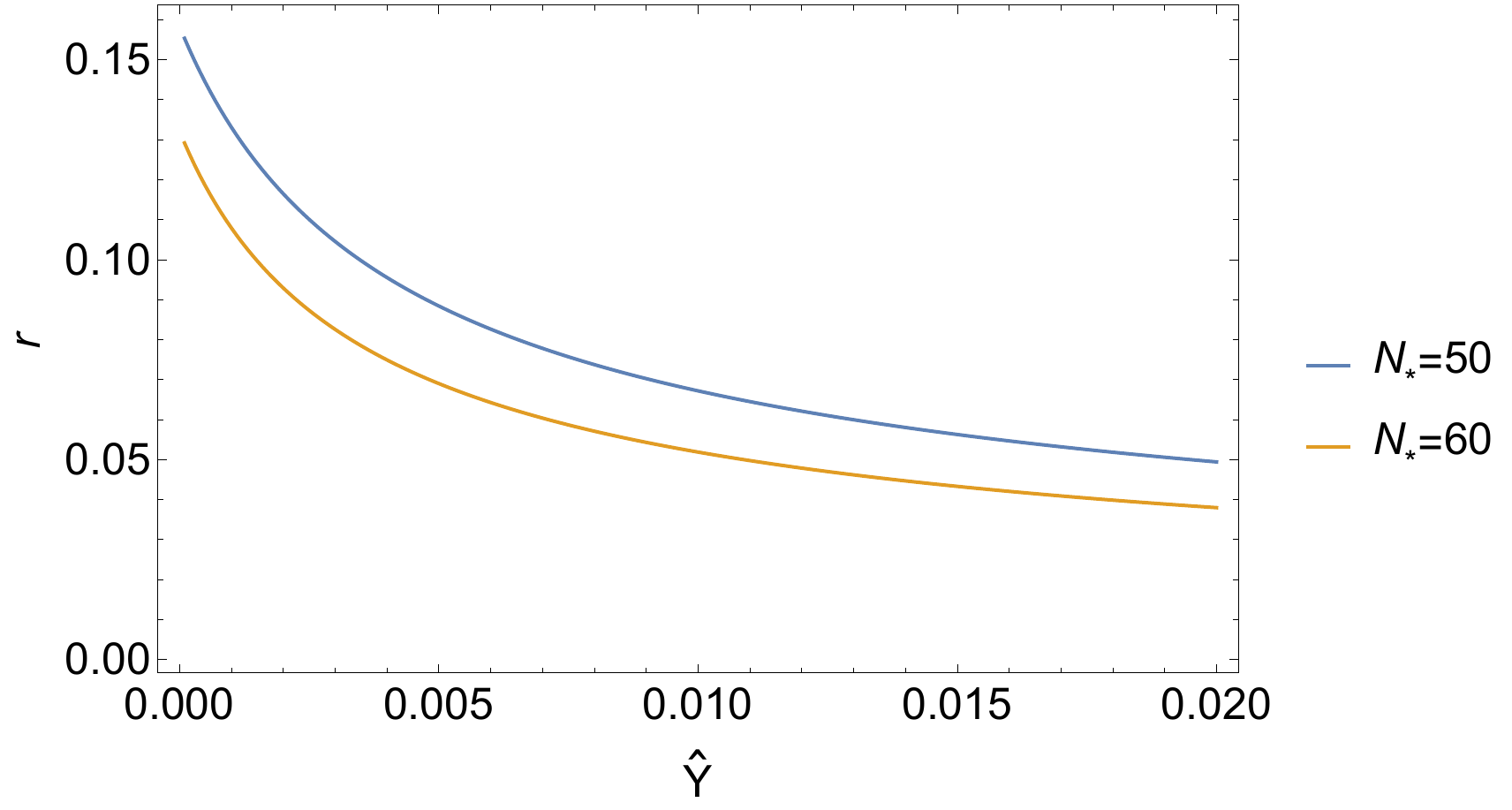}
\vspace{-.75cm}
\caption{Spectral index and tensor-to-scalar ratio in terms of $10^{-4} \leqslant \hat{\Upsilon} \leqslant 10^{-2}$  for $N_*=50$ and $N_* = 60$ e-folds.}\label{fig0}
\end{figure}
\end{center}

\begin{figure}[H]
	\begin{center}
\includegraphics[width=140mm]{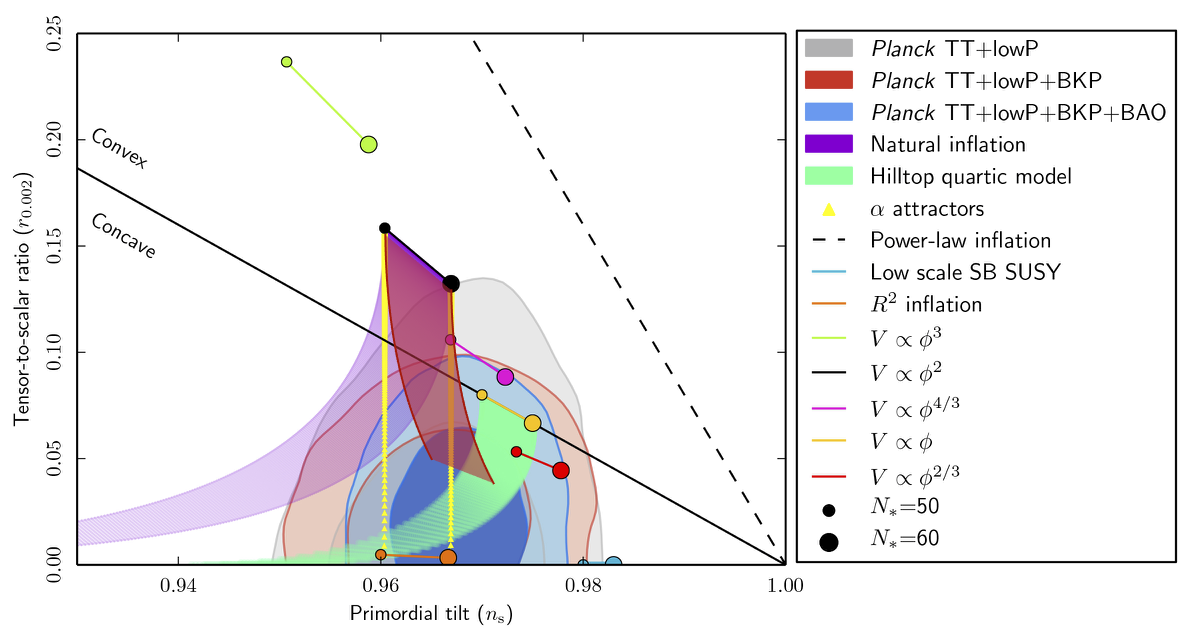}
\caption{Spectral index $n_s$ vs tensor-to-scalar ratio $r$ superimposed over the plot given by the Planck collaboration \cite{Ade:2015lrj} for the single field model with $10^{-4} \leqslant \hat{\Upsilon} \leqslant 10^{-2}$.}\label{fighat}
\end{center}
\end{figure}

Since an overall factor $V_0$ drops out in the computation of $\epsilon$ and $\eta$, in the single field limit there are only two relevant parameters in the D7-brane potential, namely $\hat G$ and $\Upsilon$. Moreover, after we add the input from the moduli stabilisation scheme of section \ref{ssec:moduli} the potential simplifies to (\ref{hatpot}) whose only relevant parameter is $\hat \Upsilon \equiv (\Upsilon -1 ) \hat G$, with typical range $10^{-4} \leqslant \hat{\Upsilon} \leqslant 10^{-2}$. We have scanned over this range of $\hat \Upsilon$ showing how the cosmological observables evolve when this parameter is varied, displaying the results in figure \ref{fig0}. We find that the typical range for these cosmological observables is $n_s  \simeq  0.96 -0.97$ and $r  \simeq 0.04 - 0.14$. 
%
%
In figure \ref{fighat} we have superimposed the precise region  in the $n_s - r$ plane over the Planck collaboration results \cite{Ade:2015lrj}.
%
%
%
\vspace{-.75cm}
%
\vspace{-.75cm}

\section{Embedding into type IIB/F-theory}
\label{sec:embedding}

Let us now consider how to construct compactifications in which the above flux-flattened 7-brane scalar potential drives large-field inflation. One important ingredient when building models of large field inflation is to provide a configuration in which the inflaton candidate is allowed to perform trans-Planckian excursions. In the case of D7-brane position moduli, this requires using the framework of F-term axion monodromy \cite{Marchesano:2014mla}, and in particular D7-branes with periodic directions in their moduli space, as already pointed out in \cite{Hebecker:2014eua,Arends:2014qca,Ibanez:2014kia,Ibanez:2014swa}. We will discuss the general features of these constructions and the relation to the D7-brane potential discussed in the previous section, paying special attention to the case of D7-branes on $\T^4/\IZ_2 \times \T^2$ and its F-theory lift to ${\bf K3} \times {\bf K3}$ \cite{Dasgupta:1999ss,Gorlich:2004qm,Lust:2005bd,Braun:2008ua,Braun:2008pz,Hebecker:2014eua,Arends:2014qca}. This simple embedding not only contains the main features of an inflationary model of mobile D7-branes, but it is also well-understood in terms of the K\"ahler and superpotential that describe the full 4d scalar potential at small field values. The latter will be crucial to understand how to generate mass hierarchies between the inflaton sector and the rest of the scalars of the compactification and, ultimately, to embed the 7-brane scalar potential into a consistent framework of moduli stabilisation, along the lines of \cite{Bielleman:2016olv}.

\subsection{Periodic 7-branes and model building}
\label{ss:periodic}

Let us consider type IIB string theory compactified in a Calabi-Yau orientifold $X_6$, and a D7-brane wrapping a holomorphic four-cycle $\CS$ in it. The moduli space of such four-cycle will depend on its topology, and in particular on the Hodge number $h^{2,0}(\CS)$ that gives the complex dimension of holomorphic deformations of $\CS$. As we are interested in mobile D7-branes, we will assume that $h^{2,0}(\CS) > 0$. The infinitesimal holomorphic deformations of $\CS$ are  given by a set of normal holomorphic vectors $\{X^i\}$ such that 
\be
\iota_{X^i} \Omega|_{\CS} \, =\, \tilde{\a}_i
\label{5chain}
\ee
where $\Omega$ is the holomorphic three-form in $X_6$ and $\tilde{\a}_i$ is a basis of (2,0)-forms in $\CS$. We may choose the $X^i$ such that the $\tilde{\a}_i$ have a constant norm, and integrate the infinitesimal deformations to define D7-brane position coordinates in terms of the chain integrals
\be
\Phi^i\, =\, \frac{1}{l_s^5} \int_{\Sig_5}  \Omega \wedge \a^i
\ee
where $\Sig_5$ is a five-chain connecting the initial four-cycle $\CS$ to a homotopic divisor $\CS'$, and $\a^i$ is a dual basis of (0,2)-forms such that $\int_{\CS} \tilde{\a}_i \wedge \a^j = \delta_i^j$, extended to $\Sig_5$. Finally, we will assume that there are one or more periodic directions in the moduli space of $\CS$, and  dub a D7-brane wrapping such a four-cycle as a periodic D7-brane.\footnote{One particular example could be a D7-brane wrapping a {\bf K3} submanifold fibered over a Riemann surface. As we will see below, this condition of periodicity can be relaxed in the more general context of F-theory compactifications.}

Let us now consider the presence of background three-form fluxes $F_3$ and $H_3$ threading $X_6$. In order to cancel the Freed-Witten anomaly \cite{Freed:1999vc,Maldacena:2001xj}, we must require that the pull-back of $H_3$ on $\CS$ vanishes in cohomology. Such a condition is trivially satisfied whenever $h^{1,0} (\CS)=0$, but in general we may have that $H_3|_{\CS}$ does not vanish identically. For simplicity let us first assume that $H_3$ is transverse to $\CS$ and so $H_3|_{\CS}=0$, as implicitly taken in the computation of the previous section, namely in (\ref{G3s2}). Then the gauge invariant worldvolume flux $\CF = \sig F - B$ is closed, and can always be taken to be harmonic in $\CS$ as this choice minimises the energy of the D7-brane. Finally, let us assume that the embedding of $\CS$ is such that at this locus the D7-brane is BPS. In practice this means that $\CF$, if non-vanishing, is a primitive (1,1)-form of $\CS$.

We may now consider deforming $\CS$ along one of its periodic directions. Here there are several possibilities depending on the topology of $\CS$. If $h^{1,1} (\CS)=1$, then there is only one harmonic (1,1)-form on $\CS$, which is necessarily its K\"ahler form and therefore non-primitive. Using the assumption that $H_3|_{\CS_4}=0$ and that the D7-brane is BPS, this means that $\CF$ must vanish on $\CS$. Now, as the D7-brane moves in its moduli space, a non-vanishing B-field and hence a flux $\CF$ will be induced in its worldvolume. Because $H$ is primitive in $X_6$ the induced B-field will be primitive in $\CS$ \cite{Gomis:2005wc}, and so $\CF$ can only be a harmonic $(2,0)+(0,2)$-form. As a result $\CF$ will be anti-self-dual, the function $f(\CF)$ will be a perfect square as in (\ref{square}) and we will recover a potential of the form (\ref{potasd}). Therefore, under the above conditions we obtain a setup similar to that in \cite{Ibanez:2014swa}, with the differences that we only have one D7-brane and no orbifold projection is present. Moreover, the potential $V(\CF)$ and kinetic function $g(\CF)$ do not need to be quadratic in $\Phi$, as the induced B-field is such that 
\be
B^{(0,2)}\, =\, c_i\, \a^i
\ee
with $c_i$ more general than a linear function of $\Phi$ and $\overline{\Phi}$. What such a B-field needs to satisfy is that, upon closing a loop in the moduli space of the D7-brane, the change in $B$ should be quantised. Hence this variation can be compensated in $\CF$ by a discrete change in $F$ and the multi-branched structure of axion-monodromy models arises. Due to that, along a closed loop $c^i$ will depend on the D7-brane position as a superposition of a linear plus a periodic function, a dependence that will be translated into the function $f(\CF)$. 

Let us now consider the case where $h^{1,1} (\CS)>1$, while still assuming that $H_3|_{\CS} = 0$ along its moduli space. Then the induced B-field will be harmonic but it may have both anti-self-dual $(2,0)+(0,2)$ and self-dual (1,1)-primitive components, depending on the components of $\iota_{X}\im\, G_3|_{\CS}$. The former will contribute to the kinetic term and potential as the quantity ${\cal G}$ in (\ref{gPhi}) and (\ref{VPhi}), while the latter will contribute as ${\cal H}$. Again, these quantities need not be the square of a linear function of $\Phi$ and $\overline{\Phi}$ as in the previous section, but rather of a linear plus a periodic function along each periodic coordinate of the D7-brane, giving a quadratic potential with modulations. In any event the potential and kinetic term will be of this form and so the effect of flux flattening will occur for large values of $\Phi$, specially when the induced B-field has an amount of self-dual component which is comparable or bigger than that of the anti-self-dual component. 

Finally, let us consider the case where $H_3|_{\CS} \neq 0$. Then, even at its BPS locus, the D7-brane will have a non-closed, co-exact induced B-field component $B^{\rm co}$ that solves $dB^{\rm co} = H_3|_{\CS}$. Now, in order to minimise the D7-brane energy, the system can always develop an exact piece for $F$, $F^{\rm ex} = d a$ such that $\CF - \CF^{\rm h} = \sig F^{\rm ex} - B^{\rm co}$ is self-dual, independently of what the harmonic component $\CF^{\rm h}$ of the worldvolume flux is. As a result, this non-closed B-field will contribute to the D7-brane potential and kinetic term as ${\cal H}$ in (\ref{gPhi}) and (\ref{VPhi}), inducing the effect of flux-flattening even in the case where $h^{1,1} (\CS)=1$. Notice however that this self-dual, non-harmonic component of $\CF$ is by definition periodic upon completing a loop in the D7-brane position space, so in order to induce a parametrically large flux flattening we need to consider the case where $h^{1,1} (\CS)>1$. 

Part of this dynamics will be captured by the 4d effective action of the compactification. In particular in the absence of fluxes we have that the K\"ahler potential capturing the 4d axio-dilaton $S$, the complex structure moduli and D7-brane kinetic terms has the form \cite{Grimm:2004uq,Jockers:2004yj,Jockers:2005zy}
\be
K = -\log\left[-\frac{i}{l_s^6} \int_{X_6} \Omega\wedge \overline \Omega\right]-\log \left[-i (S - \overline S+ \mathcal C (\Phi, \overline \Phi) )\right]\,.
\ee
where ${\cal C}$ is a real function of the D7-brane position and the complex structure moduli. Clearly, ${\cal C}$ must respect the periodicity of the moduli space of periodic D7-branes \cite{Arends:2014qca}. This will manifest as discrete shift symmetries that should be respected even when one-loop \cite{Berg:2004ek,Berg:2005ja,Haack:2008yb,Berg:2011ij,Berg:2014ama} and warping effects \cite{Shiu:2008ry,Martucci:2014ska,Martucci:2016pzt} are taken into account.

When including background and worldvolume fluxes a potential will be generated for the dilaton, complex structure and D7-brane position moduli. For small values of these fields such potential will be captured by the effective superpotential \cite{Gukov:1999ya,Jockers:2005zy,Martucci:2006ij}
\be
W = W_{\text{GVW}} + W_{D7} = \frac{1}{l_s^6} \int_{X_6} G_3 \wedge \Omega + \frac{1}{l_s^5} \int_{\Sigma_5} \Omega \wedge \mathcal F \,,
\ee
where $\Sigma_5$ is defined as in (\ref{5chain}). 

Finally, we may also understand this effective theory from the perspective of F-theory, where all the above moduli become complex structure moduli of the Calabi-Yau fourfold $Y_8$. In this case it is straightforward to write K\"ahler potential and superpotential for these moduli as \cite{Gukov:1999ya,Giddings:2001yu,Grimm:2010ks}
\be
K = -\log\left[ \frac{1}{l_M^8}\int_{Y_8} \Omega_4 \wedge \overline \Omega_4 \right]\,,
\label{KF}
\ee
\be
W = \frac{1}{l_M^8}\int_{Y_8} G_4 \wedge \Omega_4\,.
\label{WF}
\ee
As we will discuss below, this description allows to generalise the setup with a periodic D7-branes to more general compactifications in which models of F-term axion monodromy can also be constructed.

\subsection{A simple $\mathbf{K3} \times \mathbf{K3}$ embedding}

As pointed out in \cite{Hebecker:2014eua}, one simple case where periodic D7-branes are realised is in type IIB string theory compactified in an orientifold of $\mathbf {T}^4/\mathbb Z_2  \times  \mathbf T^2$, which is the orbifold limit of the ${\bf K3} \times \mathbf T^2$ orientifold. This compactification space is constructed by first considering the orbifold $\mathbf {T}^4/\mathbb Z_2  \times  \mathbf T^2$, with the $\IZ_2$ action generated by  $\theta : (z_1,z_2,z_3) \raw (-z_1,-z_2, z_3)$, and with the coordinate $z_i$ spanning the $i$-th torus. One then mods out by the orientifold action $\Omega \mathcal R (-1)^{F_L}$ with $\mathcal R : (z_1,z_2,z_3) \raw (-z_1,-z_2,- z_3)$, which introduces a total of 64 O3-plane located at the fixed loci of $\mathcal R$ as well as 4 orientifold O7-planes located at the fixed loci of $\mathcal R \cdot \theta$. In the case where no exotic O3-planes are present, the condition of cancellation of D3-brane tadpoles is 
\al{\label{eq:TadD3}N_{D3} + \frac{1}{2 l_s^4}\int_{X_6} H_3 \wedge F_3 = 16
}
where $l_s^2 = 2 \pi \sig$. Here the closed string fluxes $F_3$, $H_3$ are constant and obey the following quantisation conditions\footnote{Flux quanta should be multiples of 2 in the particular orbifold we are considering, see \cite{Frey:2002hf}.}
\al{
\frac{1}{l_s^2} \int_{\gamma_3} F_3 \in 2 \mathbb Z\,, \quad \frac{1}{l_s^2} \int_{\gamma_3} H_3 \in 2 \mathbb Z
}
for all $\gamma_3 \in H_3(X,\mathbb Z)$. Finally, cancellation of D7-brane tadpoles is ensured by introducing 16 D7-branes wrapping $\mathbf {T}^4/\IZ_2$ and being point-like in the transverse coordinates of $\mathbf T^2$, which is parametrised by the complex position field $\Phi$. Any of these D7-branes is then a periodic D7-brane with one complex modulus and two periodic directions.

One nice feature of this system is that it admits a simple embedding in a F-theory compactification on a Calabi-Yau fourfold $Y_8$ given by ${\bf K3 \times \widetilde {K3}}$. Indeed, if ${\bf \widetilde{K3}}$ is elliptically fibered upon taking the weak coupling limit we obtain a type IIB compactification on ${\bf K3}\times \mathbf \T^2$ with 16 D7-branes located at points on the torus and 4 O7-planes. Our initial setup may be easily recovered upon taking the limit in complex structure moduli space where the {\bf K3} becomes the orbifold $\mathbf T^4 /\mathbb Z_2$. The F-theory description has the advantage of describing on the same ground closed and open string moduli. Note that in this setup the cancellation of D3-brane tadpole translates to 
\al{ N_{D3} + \frac{1}{2 l^6_M} \int_{Y_8} G_4 \wedge G_4 = \frac{\chi(Y)}{24}\,,
}
where $l_M$ is the M-theory Planck length and in the case at hand $\chi(Y) = 24^2$. In this case the closed string flux $G_4$ will be quantised as\footnote{Note that for the case of ${\bf K3 \times \widetilde {K3}}$ the second Chern class satisfies $\frac{1}{2} c_2({\bf K3 \times \widetilde {K3}})\in H^4 ({\bf K3 \times \widetilde {K3}},\mathbb Z)$ and therefore the fluxes should be simply integrally quantised \cite{Witten:1996md}.}
\al{ \frac{1}{l^3_M} \int_{\gamma_4} G_4 \in \mathbb Z\,,
}
for all $\gamma_4 \in H_4(Y,\mathbb Z)$.

This F-theory description also has the advantage that provides a simple description of the 4d $\mathcal N=1$ effective action for small field values, and in particular explicit expressions for the tree-level K\"ahler and superpotentials (\ref{KF}) and (\ref{WF}), see e.g. \cite{Lust:2005bd,Arends:2014qca}. Since the holomorphic 4-form decomposes into the wedge product of the holomorphic 2-forms of each ${\bf K3}$ surface as $\Omega_4 = \Omega_2 \wedge \widetilde \Omega_2$, to express the K\"ahler potential it is convenient to introduce the period vectors $\Pi$ and $\widetilde \Pi$, respectively defined as the integrals of $\Omega_2$ and $\widetilde \Omega_2$ over a basis of integral 2-cycles. The periods of each {\bf K3} may be written as\cite{Lust:2005bd,Braun:2008ua,Braun:2008pz,Hebecker:2014eua,Arends:2014qca}
\al{\Pi = \frac{1}{2}\left(\begin{array}{c}1 \\ C^2 - \tau_1 \tau_2 \\ \tau_1 \\ \tau_2 \\ 2 C^a \end{array}\right)\,,\quad \widetilde \Pi = \frac{1}{2}\left(\begin{array}{c}1 \\ \Phi^2 - S \tau_3 \\ S \\ \tau_3 \\ 2 \Phi^a \end{array}\right)\,,
}
where $a = 1,\dots ,16$ and $C^2$ is the square of the vector $C^a$ and similarly for $\Phi^2$. When comparing with the type IIB setting we may identify the moduli $\tau_i$ with the complex structure modulus of the $i$-th torus, $S$ with the axio-dilaton, $\Phi^a$ with the relative position of the D7-branes with respect to the O7-planes and the moduli $C^a$ are the additional complex structure moduli of the first {\bf K3} surface. Using the period vectors it is straightforward to write down the K\"ahler potential (\ref{KF}) as
\al{K = - \log\big[ 2 {\overline\Pi} . M . \Pi\big] - \log\big[ 2 \overline{\widetilde\Pi}. M . \widetilde\Pi\big]\,, \label{Kperiod}
}
where $M$ is the intersection matrix
\be
M = \left(\begin{array}{ccccc} 0 & 2 & & & \\
2 & 0 & & & \\
 & & 0& 2& \\
  & & 2& 0& \\
    & & & & {\bf 1}_{16}\\
\end{array}\right)\,.
\ee
For simplicity we may take the limit where the first {\bf K3} becomes the orbifold $\mathbf T^4/\mathbb Z_2$, turning off the moduli $C^a$, and also turn off all $\Phi^a$ except one, considering a single moving D7-brane whose position is given by $\Phi$. Then we obtain that the K\"ahler potential is 
\al{K = - \log \left[- (\tau_1 -\overline \tau_1)((\tau_2 -\overline \tau_2)\right]- \log\left[-(S-\overline S)(\tau_3-\overline \tau_3)+ (\Phi-\overline \Phi)^2\right] \label{Kfields}
}
This K\"ahler potential can also be written in the form (\ref{Kperiod}) using the simplified period vectors and intersection matrix
\be
\Pi = \left(\begin{array}{c}1 \\ - \tau_1 \tau_2 \\ \tau_1 \\ \tau_2 \\ 0 \end{array}\right)\,,\quad 
\tilde \Pi = \left(\begin{array}{c}1 \\ \Phi^2- S\tau_3  \\ S \\\tau_3 \\ 2 \Phi  \end{array}\right)\,, \quad 
M = \left(\begin{array}{ccccc} 0 & 2 & & & \\
2 & 0 & & & \\
 & & 0& 2& \\
  & & 2& 0& \\
    & & & & 1\\
\end{array}\right)\,. \label{redperiods}
\ee
Finally, in this reduced moduli space, the most general superpotential (\ref{WF}) can be written as 
\be\label{eq:supP}
l_s W = \Pi . G.\widetilde \Pi\,,
\ee
where $\Pi$, $\widetilde \Pi$ are as in (\ref{redperiods}) and $G$ is a matrix of integer entries containing the relevant flux quanta
\be
G = \left(
\begin{array}{ccccc}
 \hat{n}_0 & {m}_0 & -n_0 & \hat{m}_0 & f_0 \\
 \hat{n_3} & {m}_3 & -{n}_3 & \hat{m}_3 & f_3 \\
 \hat{n_1} & {m}_1 & -{n}_1 & \hat{m}_1 & f_1 \\
 \hat{n_2} & {m}_2 & -{n}_2 & \hat{m}_2 & f_2 \\
 0 & 0 & 0 & 0 & 0 \\
\end{array}
\right)\,,
\ee
In the type IIB limit $\hat{m}_i, \hat{n}_i \in \IZ$ can be identified with quanta of $F_3$, then ${m}_i, {n}_i \in \IZ$ with quanta of $H_3$, and $f_i\in \IZ$ with D7-brane worldvolume flux quanta \cite{Arends:2014qca}. By explicit computation one finds that the superpotential reads 
\be
l_s W = \hat{\mathfrak{n}} + \hat{\mathfrak{m}} \, \tau_3 -{\mathfrak{n}}\, S + {\mathfrak{m}} \left(\Phi^2-S \tau_3\right)+ 2\mathfrak{f}\, \Phi   \,.
\label{Weff}
\ee
where the calligraphic letters are functions of the moduli of the first ${\bf K3}$, namely
%
\bea
\hat{\mathfrak{n}} & = & \hat{n}_0 + \hat{n}_1 \tau_1 + \hat{n}_2 \tau_2 - \hat{n}_3 \tau_1\tau_2\\
\hat{\mathfrak{m}} & = & \hat{m}_0 + \hat{m}_1 \tau_1 + \hat{m}_2 \tau_2 - \hat{m}_3 \tau_1\tau_2\\
\mathfrak{n} & = & n_0 + n_1 \tau_1 + n_2 \tau_2 - n_3 \tau_1\tau_2\\
\mathfrak{m} & = & m_0 + m_1 \tau_1 + m_2 \tau_2 - m_3 \tau_1\tau_2\\ 
\mathfrak{f} & = & f_0 + f_1 \tau_1 + f_2 \tau_2 - f_3 \tau_1\tau_2
\eea

As stressed above, using these K\"ahler and superpotential to compute the scalar potential for closed and open string moduli is only a good approximation in the regime of small field values for $S$, $\tau_i$ and $\Phi$. Nevertheless, these supergravity quantities are quite useful to detect discrete and continuous symmetries of our system, as we will discuss in the following. Finally, the above K\"ahler potential will be subject to one-loop corrections, see \cite{Berg:2004ek,Berg:2005ja,Haack:2008yb} for details. For simplicity, in the following we will assume that such one-loop effects are negligible. 

\subsection{Monodromies and shift symmetries}\label{mono}

\subsubsection*{Discrete symmetries and multi-branched structure}

Besides providing simple expressions for the effective K\"ahler and superpotential, the example of ${\bf K3 \times \widetilde {K3}}$ is useful in the sense that the discrete shift symmetries characteristic of axion-monodromy systems can be easily detected. Indeed, recall form the discussion of section \ref{ss:periodic} that in any type IIB flux compactification with periodic D7-branes a multi-branched potential is expected to appear, in which closing a loop in the D7-brane moduli space is compensated by shifting some worldvolume flux quanta, and that this operation corresponds to a change in the branch of the 4d potential. Such symmetry is manifest in the DBI computation of section \ref{sec:flatflux}, since the potential and kinetic terms only depend on $\CF$. When embedded in the toroidal model $\mathbf {T}^4/\mathbb Z_2  \times  \mathbf T^2$, this discrete symmetry corresponds to shifting $\Phi$ by the lattice $\Lam =  \{p + q \tau_3\}$ that describes the non-trivial loops of the ${\bf T^2}$ transverse to the D7-brane. Clearly, one would expect that such a discrete symmetry is also manifest in the 4d effective theory that arises from the ${\bf K3 \times \widetilde {K3}}$ F-theory lift of this compactification. 

In particular, one would expect that the K\"ahler potential (\ref{Kfields}) is invariant per se, as in the absence of fluxes the theory is fully symmetric under lattice shifts of $\Phi$. Indeed one sees that this K\"ahler potential is invariant under the transformations
\begin{align}
& {\rm (a)}  \ \ \ \ \,  \Phi\ \raw\ \Phi + 1\,, \label{eq:tr1}\\
& {\rm (b)}  \left\{\begin{array}{l}\Phi\ \raw\ \Phi + \tau_3\\
S\ \raw\ S + 2 \Phi + \tau_3 \,
\end{array}\right. \label{eq:tr2}\,
\end{align}
that generate the lattice $\Lam$ describing  ${\bf T^2} = \IR^2/\Lam$, and in general under the transformation
\be
\left\{\begin{array}{l}
\Phi \raw \Phi + p + q \tau_3 \\
S \raw S + 2 q \Phi + q \left( p + q \tau_3\right) \,
\end{array}\right. \quad {\rm with } \quad  p, q \in \IZ\, .
\label{lamshift}
\ee
This discrete symmetry is easier to detect in the matrix formulation of the K\"ahler potential (\ref{Kperiod}), as these transformations can be expressed as shifts of the period vector $\widetilde{\Pi}$ 
\be
\widetilde\Pi \, \raw \, \CS . \widetilde\Pi
\label{Shift}
\ee
where for $\widetilde\Pi$ as in (\ref{redperiods}) and in the case of the lattice generators we have that
\be\label{eq:mon}
\mathcal S_a = \left(
\begin{array}{ccccc}
 1 & 0 & 0 & 0 & 0 \\
 1 & 1 & 0 & 0 & 1 \\
 0 & 0 & 1 & 0 & 0 \\
 0 & 0 & 0 & 1 & 0 \\
 2 & 0 & 0 & 0 & 1 \\
\end{array}
\right) \,,
 \quad \quad
\mathcal S_b = \left(
\begin{array}{ccccc}
 1 & 0 & 0 & 0 & 0 \\
 0 & 1 & 0 & 0 & 0 \\
 0 & 0 & 1 & 1 & 1 \\
 0 & 0 & 0 & 1 & 0 \\
 0 & 0 & 0 & 2 & 1 \\
\end{array}
\right) \,,
\ee
for (\ref{eq:tr1}) and (\ref{eq:tr2}), respectively. Then because 
\be
\mathcal S^T . M . \mathcal S = M
\ee
we have that each of these shifts as well as any sequence of them leaves the K\"ahler potential invariant. 

With respect to the superpotential, we expect that the discrete symmetry is preserved if combined with discrete shifts of the flux quanta. More precisely the shift (\ref{Shift}) will be compensated by the opposite shift in the flux matrix
\be
G\, \raw \, G . \CS^{-1}
\label{Gshift}
\ee
which in the case of the lattice generator (\ref{eq:tr1}) translates into
\be
f_i\, \raw \, f_i - {m}_i\, , \quad \quad \hat{n}_i \, \raw \, \hat{n}_i+ {m}_i - 2f_i \, ,
\ee
and in the case of the generator (\ref{eq:tr2}) it becomes
\be
f_i\, \raw \, f_i + {n}_i\, , \quad \quad \hat{m}_i \, \raw \, \hat{m}_i - {n}_i - 2f_i \, .
\ee
While these discrete symmetries are derived in the context of F-theory, they have an intuitive interpretation in terms of their type IIB limit. On the one hand, the shift in $f_i$ corresponds to the shift in D7-brane worldvolume flux quanta that compensates the shift of B-field, as discussed in section \ref{ss:periodic}. On the other hand, the shifts in $\hat{n}_i$, $\hat{m}_i$ correspond to shifts in the background flux $F_3$ due to the rearrangement of D5-brane charge. 

This example allows us to readily generalise the picture of discrete shift symmetries to a generic Calabi-Yau four-fold. Here the fundamental quantity is the period vector $\Pi(z)$ of the 
Calabi-Yau four-fold whose entries are functions of the four-fold complex structure moduli. Notice that since in F-theory brane position moduli get unified with closed string moduli we can treat them on equal footing.
In this scenario the tree-level K\"ahler potential is written as
\al{K = - \log \left[\overline \Pi . M . \Pi\right]\,,
}
where $M$ is the intersection matrix of integral 4-cycles in the Calabi-Yau four-fold. A discrete shift symmetry is present whenever upon performing a suitable translation in
complex structure moduli space $z \raw z + f(z)$ it is possible to find a matrix $\mathcal S$ with integer coefficients such that $\Pi(z+ f(z)) = \mathcal S. \Pi(z)$ and 
$\mathcal S^T . M . \mathcal S = M$. While this clearly constitutes a symmetry of the K\"ahler potential it is necessary to take into account how the superpotential transforms as well
if fluxes are added. The superpotential may be easily expressed in terms of the period vector as
\al{l_s W = \mathsf{G}. M . \Pi(z)\,,
}
where $\mathsf{G}$ is a vector with integer coefficients. Upon performing the aforementioned discrete transformation we find that the transformed superpotential is
\al{l_s W ' = \mathsf G' . M . \Pi(z) \,,
}
where $\mathsf G' = \mathsf G. (\mathcal S^T)^{-1}$. This shows how the effect of performing a discrete shift symmetry is translated in a suitable redefinition of the integer flux
quanta, a mechanism which is the avatar of axion monodromy. It is important to state that the presence of these discrete shift symmetries effectively cuts the moduli space to
some \emph{fundamental domain} which may contain some compact directions inside it: addition of fluxes effectively unfolds this compact moduli space, a signature of axion 
monodromy. Identification of the correct fundamental domain is in general case is a difficult exercise although in some specific cases the answer is known \cite{Candelas:1990rm,Klemm:1992tx,Conlon:2016aea}.

The question that remains open is when and under which conditions a discrete shift symmetry does appear. Luckily it is possible to find an answer to these questions: 
discrete shift symmetries are intimately tied with the presence of singular points in the complex structure moduli space.\footnote{In some cases though the presence of a singular point 
in the complex structure moduli space does not give discrete shift symmetries, see \cite{Garcia-Etxebarria:2014wla} for examples.} In the case we have previously analysed the singularity is located at the point of large complex structure of the Calabi-Yau 4-fold, and indeed in the proximity of this point a shift symmetry appears for the complex structure moduli \cite{Arends:2014qca}.
For simplicity we will phrase our discussion in the case of complex structure moduli space of a Calabi-Yau 3-fold $X_6$, where most examples are known, although the discussion can be easily generalised to a Calabi-Yau $n$-fold mutatis mutandis. First we need to highlight one of the characteristics of the period vector $\Pi(z)$: namely that it behaves as a section of an appropriate vector bundle $\mathcal H$ over the complex structure moduli space $\mathcal M$. 
Specifically at $z\in \mathcal M$ the fibre of $\mathcal H$ is simply $H^3(X_z,\mathbb Z)$ where $X_z$ is the Calabi-Yau manifold $X$ with complex structure specified by $z$.
This vector bundle comes equipped with a flat connection $\nabla$ called Gau$\ss$-Manin connection which allows to perform parallel transport of sections of $\mathcal H$
around paths on $\mathcal M$. While it is true that the connection is flat (and therefore parallel transport around closed cycles would give no transformations on sections of $\mathcal H$), it may develop some singularities at specific points in the complex structure moduli space $\hat z_i$ where the Calabi-Yau manifold develops a singularity. The presence of singularities in the Gau$\ss$-Manin connection implies that upon circling these singular points a section of $\mathcal H$ gets acted upon by a matrix transformation which realises the transformation of the period vector $\Pi(z)$ advocated above. This provides a mechanism to realise discrete shift symmetries in general Calabi-Yau compactifications, although the precise details of the vector period transformations are somewhat technical and here we will refrain from delving into them. The interested reader may consult for instance \cite{Greene:2000ci,Eguchi:2005eh,Donagi:2012ts,Anderson:2013rka} and references therein for explicit examples. 

\subsubsection*{Continuous shift symmetries}

One well-known fact is that in the tree-level K\"ahler potential (\ref{Kfields}) the discrete shift symmetry (\ref{eq:tr1}) is promoted to the continuous shift-symmetry
\be
 \Phi\ \raw\ \Phi + \lam 
 \label{tshift}
\ee
with $\lam \in \IR$. This continuous symmetry highlights the field direction $\re\, \Phi$, and makes it a natural inflaton candidate, as considered in \cite{Bielleman:2016olv}. 

While (\ref{tshift}) is an obvious shift symmetry of this K\"ahler potential it is strange that it is the only one. After all, it is nothing but a translation along one of the one-cycles of the ${\bf T}^2$ transverse to the D7-brane. Geometrically all of these one-cycles are on the same footing, and microscopically they are all similar for the D7-brane. Hence there is a priori no reason why the field direction (\ref{tshift}) should be special. In particular we would expect to find a continuous shift symmetry like (\ref{tshift}) for each of the points of the lattice that defines ${\bf T}^2$. 

One can indeed see that this is the case whenever we allow for field space excursions involving $S$ and $\Phi$ simultaneously. Indeed, let us consider our ${\bf K3} \times {\bf \widetilde{K3}}$ model with an initial point in moduli space given by $(\Phi_0, S_0)$ and with all  $\tau_i$ fixed to some value. Then if we consider the one-dimensional trajectory
\be
\left\{\begin{array}{lcl}
\Phi & = & \Phi_0 + \lam \left( s + r \tau_3 \right) \\
S & = & S_0 + r \frac{\Phi^2-\Phi_0^2}{s + r \tau_3}  \,
\end{array}\right. \quad \text{ with varying } \lam \in \IR
\label{Kshift}
\ee
and fixed $r,s \in \IR$, one can see that the K\"ahler potential (\ref{Kfields}) is left invariant. Notice that we do not have one shift symmetry but an infinite number of them, parametrised by $(r,s) \in \IR^2$. If we take $(r,s)= (p,q) \in \IZ^2$ then each of these trajectories connects with different lattice points of ${\bf T}^2$, where they reduce to (\ref{lamshift}). In particular, taking $(r,s) = (0,1)$ and $\lam \in \IN$ we generate the discrete shifts that correspond to (\ref{eq:tr1}) and taking $(r,s) = (1,0)$ we generate those in (\ref{eq:tr2}).

We then see that, when combining field excursions involving $\Phi$ and $S$, many shift symmetries arise, and that they are related to the periodic directions in the D7-brane moduli space. Absent some criterium that selects one among the rest, they are all equally valid as inflationary trajectory candidates and should be considered on equal footing. 

The criterium to select one trajectory among all of them will in general come from the effective superpotential. Indeed, as discussed above $W$ will transform non-trivially under discrete shifts that leave $K$ invariant, and generically the same will happen for their continuous counterparts. Interestingly, for the case under discussion one can easily characterise whenever $W$ selects one of the above trajectories among the others. Indeed, it is easy to check that for a superpotential of the form (\ref{Weff}) a trajectory with fixed $\tau_i$ and
\be
\left\{\begin{array}{lcl}
\Phi & = & \Phi_0 + \kappa \\
S & = & S_0 + \mathfrak{n} \frac{\Phi^2-\Phi_0^2}{\mathfrak{m} + \mathfrak{n} \tau_3} + 2\mathfrak{f} \frac{\Phi-\Phi_0}{\mathfrak{m} + \mathfrak{n} \tau_3}  \,
\end{array}\right. \quad \text{ with varying }  \kappa \in \IC
\label{Wshift}
\ee
leaves $W$ invariant. As a result, whenever $\mathfrak{f} = 0$ and $ \mathfrak{n}\, \overline{ \mathfrak{m}} \in \IR$ there will be a field space trajectory of the form (\ref{Kshift}) that leaves both the K\"ahler and superpotential invariant, which signals a flat direction of the scalar potential. As discussed in Appendix \ref{ap:SL} this can be made manifest by using the $SL(2,\IR)$ invariance of $K$. 

As we will see in the following, this result will still hold when we complete $K$ and $W$ with the remaining ingredients to describe a compactification with full moduli stabilisation. Therefore, in such a setup we will have a simple mechanism to generate flat directions in field space, which then will be useful to generate mass hierarchies among fields in the scalar potential. 

\subsection{Moduli stabilisation}
\label{ssec:moduli}

Following \cite{Bielleman:2016olv}, one may try to embed a system with a mobile D7-brane into a type IIB compactification with the necessary ingredients for full moduli stabilisation. In the case where $h^{1,1} (\CS)=1$ and the background flux is transverse to $\CS$, one may capture the non-trivial kinetic term of the D7-brane position field  in terms of a higher derivative correction to the K\"ahler potential, as done in \cite{Bielleman:2016grv,Bielleman:2016olv}, and so study the stability of the inflationary trajectory by means of 4d supergravity techniques. In the case where the effect of flux flattening is important, namely when $h^{1,1} (\CS)>1$, such a description for the D7-brane scalar potential and kinetic terms for large values of $\Phi$ is not known. Nevertheless, one may still use 4d supergravity to analyse the stability of the inflationary trajectory at small field values, in order to estimate how important are the effects of moduli stabilisation and heavy field backreaction on the naive potential computed in section \ref{sec:flatflux}. 

\subsubsection*{Recovering the DBI potential at small field}

In order to connect with the setup of section \ref{sec:flatflux} let us assume a D7-brane whose moduli space of positions contains a $\T^2$ parametrised by the complex field $\Phi$. Then, by analogy with the ${\bf K3 \times \widetilde {K3}}$ example, we may consider that the D7-brane and closed string dynamics is governed by an effective superpotential of the form
\be
l_s W\, =\, \hat{f}- S {f} +  \left(\Phi^2 - S U \right) {g} + U \hat{g}
\label{Weffp}
\ee
where $U$ is the complex structure modulus of such a $\T^2$ and $f$, $g$, $\hat{f}$, $\hat{g}$ are holomorphic functions of the flux quanta and the complex structure moduli of the compactification. Similarly, one would expect a K\"ahler potential of the form
\be
K \, =\,  - \log\left[(\Phi-\overline \Phi)^2-(S-\overline S)(U-\overline U)\right] + K_2
\label{Keffp}
\ee
where $K_2$ contains the dependence on the K\"ahler and remaining complex structure moduli. 

In the absence of any superpotential for the K\"ahler moduli we will recover a positive definite scalar potential which, at $\Phi=0$, reduces to the no-scale scalar potential in \cite{Giddings:2001yu} for the axio-dilaton $S$ and complex structure moduli. In principle, one may assume that the mass for these fields at the vacuum is much larger than that of $\Phi$ and so, following the philosophy in \cite{Landete:2016cix}, replace such heavy fields by their vevs in (\ref{Weffp}) and (\ref{Keffp}). This strategy, followed in \cite{Ibanez:2014swa,Bielleman:2016olv}, is however only a fair approximation for a restricted range of superpotential parameters in (\ref{Weffp}). Indeed, from the discussion above we have that whenever $g/f \in \IR$ there is a flat direction of the scalar potential along $\Phi \propto f + {g} U$ in which the dilaton varies as\footnote{This assumes that in (\ref{Keffp}) $K_2$ does not depend on $S$ and $\Phi$ or, if it does, it depends through the combination $(\Phi-\overline \Phi)^2-(S-\overline S)(U-\overline U)$.}
\be
S\, =\, S_0 + \frac{g}{f}\frac{\Phi^2}{1 + \frac{g}{f}U}\, ,
\ee
with $S_0$ the vev of $S$ at $\Phi =0$. Therefore, for generic $g/f$ it is not a good approximation to assume that $S$ will remain close to its vev $S_0$.  This means that, in general, we cannot apply the philosophy of \cite{Landete:2016cix} to $S$.

Instead we can integrate out $S$ by cancelling its F-term, solving for it in terms of the other moduli and plugging the result back into the scalar potential. For simplicity, let us consider the K\"ahler and superpotential above with all the complex structure moduli including $U$ fixed to their vev. Then the F-term for $S$ is given by 
\be
D_S W \, = \, - \frac{(\Phi - \overline \Phi)^2( f +  g U) - (U-\bar U) \left( \hat{f}- \bar S {f} +  \left(\Phi^2 - \bar S U \right){g} + U \hat{g}\right)}{(\Phi - \overline \Phi)^2- (S - \bar S)(U - \bar U)}
\ee
where we have assumed that $\p_S K_2 = 0$. Hence we obtain $D_SW =0$ by demanding that 
\be
S\, =\, S_0 + \frac{\bar{g}}{\bar{f}} \frac{ \ov{\Phi}^2}{1+\frac{\bar{g}}{\bar{f}} \ov{U}}+\frac{(\Phi-\overline{\Phi})^2}{U-\ov{U}}
\label{FtermS}
\ee
Plugging this expression into the scalar potential we obtain that
\be
V \, =\, \frac{e^K}{\kappa_4^2} K^{\Phi\bar{\Phi}} |D_\Phi W|^2 \, =\, \frac{1}{4\pi \kappa_4^4} \frac{2\left|\left(\bar{f}+\bar{g}\bar{U}\right)\Phi - \left( \bar{f} + \bar{g}U\right) \bar{\Phi}\right|^2}{8 {\rm Vol}_{X_6}^2 |U - \bar{U}| |\int_{X_6} \Omega\wedge \overline{\Omega}|}  \, .
\label{Vsugraeff}
\ee 
where in our conventions $\kappa_4^2= l_s^2/4\pi$ and all volumes are measured in units of $l_s$. 

In order to compare this result with the scalar potential of section \ref{sec:flatflux} we need to canonically normalise the position field at $\Phi=0$. Taking into account that there its kinetic term is given by $K_{\Phi \overline \Phi}|_{\Phi=0} = g_s /|U-\overline U|$, with $g_s^{-1} = \im\, S_0$ we obtain that the scalar potential is
\al{V_{\text{SUGRA}} = \, \frac{g_s^{-1}}{2\pi \kappa_4^2} \,\frac{\left|\left(\bar{f}+\bar{g}\bar{U}\right)\Phi - \left( \bar{f} + \bar{g}U\right) \bar{\Phi}\right|^2}{8 {\rm Vol}_{X_6}^2|\int_{X_6} \Omega\wedge \overline{\Omega}|}}  
where now $\Phi$ is canonically normalised at the origin. We may now compare with the DBI result  (\ref{eq:scpot}) in the small field limit and in the 4d Einstein frame
\be
V_{\rm DBI+CS} \, \simeq\, \frac{g_s}{\kappa_4^4} \frac{|\overline G_{\bar 1\bar 2\bar 3}  \Phi - S_{\bar 3 \bar 3} \overline \Phi|^2}{{16 {\rm Vol}_{X_6}}}
\label{VDBIeff}
\ee
where for simplicity we have set a trivial warp factor $Z=1$. We then obtain that
\be
 G_{\bar 1\bar 2\bar 3} \, = \, \frac{\kappa_4}{\sqrt{\pi}} \frac{f + g U}{g_s  {\rm Vol}_{X_6}^{1/2} |\int_{X_6} \Omega\wedge \overline{\Omega}|^{1/2}} \quad \quad \quad  S_{\bar 3\bar 3}\, =\, \frac{\kappa_4}{\sqrt{\pi}} \frac{\bar{f} + \bar{g}U}{g_s {\rm Vol}_{X_6}^{1/2} |\int_{X_6} \Omega\wedge \overline{\Omega}|^{1/2}} 
\label{dict}
\ee
Finally, as in \cite{Ibanez:2014swa} we may diagonalise this scalar potential as 
\be
\kappa_4^4\, V_{\rm DBI+CS}  \simeq \frac{g_s}{16{\rm Vol}_{X_6}} \left[ \left(|G_{\bar 1\bar 2\bar 3}| -  |S_{\bar 3\bar 3}|\right)^2 \left(\re\, \Phi'\right)^2 + \left(|G_{\bar 1\bar 2\bar 3}| +  |S_{\bar 3\bar 3}|\right)^2 \left(\im\, \Phi'\right)^2 \right]
\label{dbipot}
\ee
where 
\be
\Phi' = e^{-i\g/2} \Phi \quad \quad \quad \g = {\rm Arg\, } \left( G_{\bar 1\bar 2\bar 3}  S_{\bar 3\bar 3}\right)\, .
\label{Phi'}
\ee
Notice that using the dictionary (\ref{dict}) we have that $g \bar{f} \in \IR$ is equivalent to $|G_{\bar 1\bar 2\bar 3}| = |S_{\bar 3\bar 3}|$, which precisely is where we obtain a flat direction in the scalar potential, in agreement with our previous discussion. Away from the flat direction condition we have that the masses of the two mass eigenstates go like
\bea
\label{massImPhi}
 m_{\sqrt{2}{\rm Im \,  } \Phi'} & =  &  \frac{ g_s^{1/2}}{2\kappa_4^2 {\rm Vol}_{X_6}^{1/2}} \left( |G_{\bar 1\bar 2\bar 3}| + |S_{\bar 3\bar 3}|\right) \, =\,     2\,  e^{K/2}| W_0| (1 + \varepsilon) \, , \\
 m_{\sqrt{2}{\rm Re \,  } \Phi'} & = &  2\, e^{K/2} |W_0|\, |\varepsilon|\, ,
\label{massRePhi}
\eea
where 
\bea
|W_0|  & = & \kappa_4^{-2} |G_{\bar 1\bar 2\bar 3}|\,  {\rm Vol}_{X_6}^{1/2} \, \left|\int_{X_6} \Omega\wedge \overline{\Omega}\right|^{1/2}  = \frac{\kappa_4^{-1}}{\sqrt{\pi}}g_s^{-1}|f +gU| \, ,\\
\varepsilon & = & \frac{|S_{\bar 3\bar 3}| - | G_{\bar 1\bar 2\bar 3}|}{2 |G_{\bar 1\bar 2\bar 3}|}\, \simeq\,  \frac{\im \, U}{|f +gU|^2}\, \im\, (g\bar{f})\, .
\label{delta}
\eea
Here $\varepsilon$ measures the departure form the flat direction case, and whenever $|\varepsilon| \ll 1$ we have that $\re\, \Phi'$ is a very light compared to $\im\, \Phi'$. In that case, the heaviest mode $\im\, \Phi'$ is in turn much lighter than the complex structure and axio-dilaton moduli whenever $\kappa_4 |W_0| \ll N$, with $N$ the typical value for the flux quanta.\footnote{For instance, for the choices $\kappa_4 W_0 \sim 0.1$, $|\varepsilon| \sim 0.01$, $e^K \sim 10^{-5}$ one recovers an inflaton mass of the order $m_{\sqrt{2}{\rm Re \,  } \Phi'}^2 \sim 4 \times 10^{-11} M_P^2$ and $m_{\sqrt{2}{\rm Im \, } \Phi'}^2 \sim 4 \times 10^{-7} M_P^2$, while $m_{\rm flux}^2 = N^2 \times 10^{-5} M_P^2$. \label{scales} } In particular, its mass will not be far from that of the K\"ahler moduli sector in standard moduli stabilisation schemes. Therefore, one should be able to describe an $\CN=1$ effective field theory for $\Phi$ and the K\"ahler moduli below the flux scale, as we discuss in the following.

\subsubsection*{Integrating out the dilaton}

As mentioned above, in general it will not be a good approximation to fix the 4d axio-dilaton $S$ at its vev $S_0$ in $K$ and $W$, since $S$ varies significantly as we change the value of $\Phi$. However, when a flat direction is developed because $g\bar{f} \in \IR$, we have that  the holomorphic field redefinition
\be
\hat{S}\, =\, S - \frac{g}{f} \frac{ {\Phi}^2}{1+\frac{g}{f} {U}}
\label{hatS}
\ee
is such that $\hat{S}$ remains constant and equal to $S_0$ along the flat direction. Therefore, for describing the scalar potential in a field space region around the flat direction trajectory, one may apply the strategy of \cite{Landete:2016cix} to this new holomorphic variable $\hat{S}$, and replace it by its vev $S_0$ both in $K$ and $W$, as done with the complex structure moduli.\footnote{Notice that $\hat{S}$ is not only holomorphic on $\Phi$ and $U$, but also on all the remaining complex structure moduli through $g$ and $f$. Therefore (\ref{hatS}) can be seen as a field redefinition even at the flux scale, and one may  apply the strategy of \cite{Landete:2016cix} to all complex structure moduli and $\hat{S}$ simultaneously. We discuss alternative definitions to the definition (\ref{hatS}) in Appendix \ref{ap:SL}.} 

Whenever the flat direction is not present because $\im \, (g\bar{f}) \neq 0$ then $\hat{S}$ will no longer be constant along the trajectory of minimum energy. On the one hand it will still be true that, if $S$ is given by (\ref{FtermS}), then $\re\, \hat{S} = \re\, S_0$ for any value of $\Phi$. On the other hand it will happen that $\im\, \hat{S}$ will depart from $\im\, S_0$ as we move away from $\Phi =0$ along the said trajectory. Nevertheless, one expects that this displacement is small as long as the mass of $\im\, \Phi'$, $\re\, \Phi'$ is much smaller than the typical mass scale induced by fluxes. In particular whenever $|\varepsilon|, \kappa_4 |W_0| \ll 1$, the approximation of taking $\hat S  = S_0$ in $K$ and $W$ should be accurate enough to describe the inflationary potential up to subleading backreaction effects \cite{Landete:2016cix}. 

Doing this procedure in the no-scale case we find an effective K\"ahler and superpotential for $\Phi$ given by 
\bea\nonumber
K & = &  -  \log \left[-(S_0 - \bar S_0)( U - \bar U)  -\left(\frac{g}{f} \frac{\Phi^2}{1+\frac{g}{f} {U}} -\frac{\bar{g}}{\bar{f}}\frac{\overline \Phi^2}{1+\frac{\bar{g}}{\bar{f}} {\bar{U}}}  \right)(U-\bar U)  + (\Phi - \overline \Phi)^2\right] + K_2\\
W & = & W_0
\label{KWeffns}
\eea
where again $K_2$ contains all the dependence on the K\"ahler moduli. In terms of the components of the field $\Phi'$ defined in (\ref{Phi'}) the first part of the K\"ahler potential $K'$ reads
\bea
K' & = &    -  \log \Big[-(S_0 - \bar S_0)( U - \bar U)   \\ \nonumber
& & \left. + \oh \left(1+ \frac{|f+g\bar{U}|}{|f+gU|} \right) (\Phi' - \overline{\Phi}')^2 -  \oh \left(1- \frac{|f+g\bar{U}|}{|f+gU|} \right)(\Phi' + \overline{\Phi}')^2\right]  \\ \nonumber
 & = & -  \log \left[-(S_0 - \bar S_0)( U - \bar U) + \oh \left(1+ \frac{|S_{\bar 3 \bar 3}|}{|G_{\bar 1\bar 2\bar 3}|} \right) (\Phi' - \overline{\Phi}')^2 -  \oh \left(1-  \frac{|S_{\bar 3 \bar 3}|}{|G_{\bar 1\bar 2\bar 3}|} \right)(\Phi' + \overline{\Phi}')^2\right]
\eea
Therefore we recover an effective theory with a constant superpotential and a K\"ahler potential with no apparent shift symmetry for any component of $\Phi$. Notice however that whenever $g\bar{f} \in \IR$ or equivalently $|G_{\bar 1\bar 2\bar 3}| = |S_{\bar 3\bar 3}|$ we recover a shift symmetry along $\re\, \Phi'$, which then becomes a flat direction. Finally, we can rewrite the K\"ahler potential in the simpler form
\be
K  =  - 3 \log (T+\bar{T}) -  \log \left[4su + (1+\varepsilon) (\Phi' - \overline{\Phi}')^2 + \varepsilon (\Phi' + \overline{\Phi}')^2\right] + K_2 
\label{Kdelta}
\ee
with $u = \im\, U$, $s = \im \, S_0$ and $\varepsilon$ is defined as in (\ref{delta}). Again, notice that in the regime of interest $|G_{\bar 1\bar 2\bar 3}| \simeq |S_{\bar 3\bar 3}|$ and so $|\varepsilon| \ll 1$.
%


\subsubsection*{Adding K\"ahler moduli stabilisation}

Let us now add the necessary ingredients to achieve full moduli stabilisation in a semi-realistic setup. Since our setup requires $|W_0| \ll 1$ in order to decouple the D7-brane position modulus from the complex structure moduli, it is more natural to consider a KKLT-like scheme with a single K\"ahler modulus $T$, as done in \cite{Bielleman:2016olv}. We then have a K\"ahler potential of the form
\be
K \, =\, - 3 \log (T+\bar{T})  - \log\left[(\Phi-\overline \Phi)^2-(S-\overline S)(U-\overline U)\right] + K' 
\label{KMS}
\ee
where $K'$ contains the dependence in the complex structure moduli besides $U$. In addition we have a superpotential of the form 
\be
l_s W\, =\, l_s W_{\rm flux} + l_s W_{\rm np} \, =\, \left(\hat{f}- S {f} +  \left(\Phi^2 - S U \right) {g} + U \hat{g}\right) + l_s A \,e^{-aT} 
\label{WMS}
\ee
where $f, g, \hat{f}, \hat{g}$ depend on the flux quanta and complex structure moduli, and so may the non-perturbative prefactor $A$.  From these two quantities we compute the supergravity scalar potential
\be
V_{\rm SUGRA}\, =\, \frac{e^K}{\kappa_4^2} \left( K^{\a\bar{\b}} D_\a W D_{\bar{\b}} \overline{W} - 3 |W|^2\right)
\ee
which together with an uplifting term\footnote{Here we are treating $\Delta^2$ as a constant, as it would arise by considering, e.g., F-term uplift. As in \cite{Bielleman:2016olv} we will not delve on the actual microscopic origin of this uplifting mechanism, as it will not affect the subsequent discussions.} 
\be
V_{\rm up} \, =\, \frac{e^K}{\kappa_4^4} \Delta^2 
\label{Vup}
\ee
give us the final scalar potential
\be
V\, =\, V_{\rm SUGRA} + V_{\rm up}\, .
\label{Vtot}
\ee

Notice that if $W_{\rm np}$ does not depend on $S$ and $\Phi$ the full superpotential will still be invariant under the complex shift (\ref{Wshift}). Hence, if we also assume that $K'$ does not depend on $S$ and $\Phi$ and follow our previous discussion, we have that whenever $g \bar{f} \in \IR$ there will be a real shift of the form
\be
\left\{\begin{array}{lcl}
\Phi & = & \Phi_0 + \lam \left( 1 + \frac{g}{f} U \right) \\
S & = & S_0 + \frac{g}{f} \frac{\Phi^2}{1 + \frac{g}{f} U}  \,
\end{array}\right. \quad \text{ with } \lam \in \IR
\label{KWshift}
\ee
that leaves $W$ and $K$ invariant. Therefore both $V_{\rm SUGRA}$ and $V_{\rm up}$ will be invariant and this direction in field space will be a flat direction of the full scalar potential.  

We may now consider relaxing the above assumptions on $W_{\rm np}$ and $K'$. For instance, let us consider a non-trivial dependence of the prefactor $A$ on $\Phi$, as done in \cite{Ruehle:2017one}. In general, such a dependence may or may not be periodic in the lattice of $\Phi$. If on the one hand it is not periodic, then it should be such that $A$ is invariant under the discrete shift symmetry of section \ref{mono} that shifts fields and flux quanta simultaneously. Therefore, it will most likely depend on $\Phi^2$ through a function of $W_{\rm flux}$, and so it will be invariant under the real shift symmetry (\ref{KWshift}). If on the other hand the dependence is periodic it must be bounded, so we expect it to be subdominant with respect the dependence in $W_{\rm flux}$ for large values of $\Phi$. The same observations apply to the potential dependence of $K'$ on $\Phi$, for instance through one-loop corrections, which as stated above we assume negligible. Therefore, up to this degree of approximation the full scalar potential should develop a flat direction whenever $g \bar{f} \in \IR$, and a very light direction in field space whenever we slightly violate this condition. In the following we will consider the consequences of this feature in the simplest case, namely when $A$ and $K'$ do not depend on $\Phi$.

As in our previous discussion of the no-scale case, the variable $\hat{S}$ defined in (\ref{hatS}) remains constant and equal to its vev along such a flat direction of $V$, and very close to it when $|\delta| \propto \im (g/f)$ is very small. We may then apply the strategy of \cite{Landete:2016cix} to $\hat{S}$ and all the complex structure moduli, replacing them by their vevs in $W$ and $K$. We thus obtain an effective potential for $T$ and $\Phi$ of the form (\ref{Vtot}), where now $V_{\rm SUGRA}$ and $V_{\rm up}$ only depend on $T$ and $\Phi$, through the quantities
\bea
\label{KWMS}
W & = & W_0 + A \,e^{-aT} \\ \nonumber
K & = &  - 3 \log (T+\bar{T}) -  \log \left[4 s u  +(1+\varepsilon) (\Phi' - \overline{\Phi}')^2 +\varepsilon (\Phi' + \overline{\Phi}')^2\right]
\eea
where $u$, $s$ and $\varepsilon$ are as in (\ref{Kdelta}). All these quantities as well as $a \in \IR$ and $A$, $W_0 \in \IC$  are treated as constants. Notice that even if the inflaton candidate $\re\, \Phi'$ appears in the K\"ahler potential there is a priori no $\eta$-problem, as $|\varepsilon| \ll 1$ and so the kinetic term for $\Phi$ is dominated by the coefficient of $\im\, \Phi'$ in $K$.

Given this effective theory, we are able to stabilise the K\"ahler modulus as in the KKLT proposal \cite{Kachru:2003aw}. Cancelling the F-term of $T$ in the vacuum we arrive to the relation
\be
D_{T}W = 0 \rightarrow  W_{0} = -\frac{1}{3} A e^{-a T_{0}} (2 a \re\, T_{0}+3) \, ,
\ee
where $T_0$ is the value of $T$ at the KKLT AdS vacuum. For simplicity, in the following we will assume that $W_0$, $A \in \IR$, so that $\im\, T_0 =0$. The introduction of the uplifting term (\ref{Vup}) will shift the K\"ahler modulus vev. For instance, in order to obtain a Minkowski vacuum state one should minimise the scalar potential for every field in the vacuum and impose $V|_{\text{tot}}^{\text{vac}} = 0$ from which we obtain the following relations 
\be
A = -\frac{3 W_{0} e^{a t} (a t -1)}{2 a^2 t^2+4 a t-3} \quad \text{ ,  } \quad \Delta^2 = \frac{12 a^2 t^2  \left(a^2 t^2+a t-2\right)}{ \left(2 a^2 t^2+4 a t-3\right)^2} W_{0}^2 \kappa_4^2 \, ,
\ee
describing implicitly the new value for $t = \langle \re\, T \rangle$, while $\langle \im\, T \rangle$ still vanishes.

We can see that the ingredients for K\"ahler moduli stabilisation do not change significantly the mass hierarchies obtained in the no-scale case. Indeed, if we denote by $\varphi$ and $\xi$ the canonically normalised components $\re\, \Phi'$ and $\im\, \Phi'$, respectively, we find that in the vacuum
\be
m_{\varphi}^{2} = \frac{\varepsilon^2 W_{0}^2}{8 us t^3} + \mathcal{O}\left(t^{-4}\right)\text{  ,  } m_{\xi}^{2} = \frac{ W_{0}^2 (1+\varepsilon)^2}{8 ust^3} +\mathcal{O}\left(t^{-4}\right)\, ,
\ee
\be
m_{{\rm Re} T}^2 = \frac{a^2 W_{0}^2}{8 ust}-\frac{5 \left(a W_{0}^2\right)}{8 ust^2 }+ \mathcal{O}\left(t^{-3}\right) \text{  ,  } m_{{\rm Im } T}^2 = \frac{a^2 W_{0}^2}{8 ust}-\frac{3 \left(a W_{0}^2\right)}{8 ust^2 }+ \mathcal{O}\left(t^{-3}\right)\, .
\ee
which reproduces (\ref{massImPhi}), (\ref{massRePhi}) and the usual mass for the K\"ahler modulus in KKLT-like schemes. Again, the mass of the inflaton candidate is strongly suppressed with respect the other moduli by the parameter $\varepsilon$, and the mass of its partner $\xi$ is of the same order of magnitude as the K\"ahler moduli sector. Multifield effects during inflation will then be negligible as long as
\be
|\varepsilon| < 10^{-2}\, . 
\ee

Given these expressions, one is able to accommodate a realistic setup by for instance taking the following set of parameter values
\be
\kappa_4 A = -1.6\,, \quad a = \frac{2 \pi}{15} \,, \quad \kappa_4 W_{0} = 0.09 \,, \quad  s u= 10 \,, \quad\varepsilon = 2.3 \times 10^{-2} \, ,
\label{parameters}
\ee
so that the Minkowski vacuum is found for
\be
t = 10.8 \text{  ,  } \Delta^2 = 0.0148\, 
\ee
and the above masses are given by 
\be
m_{\varphi} = 6.4 \times 10^{-6} M_{P} \text{  ,  }  m_{\xi} = 2.8 \times 10^{-4}M_{P} \text{  ,  }  m_{{\rm Re} T} = 8.1 \times 10^{-4}M_{P} \text{  ,  }  m_{{\rm Im} T} = 9.9 \times 10^{-4}M_{P} \, .
\label{benchmasses}
\ee

\subsubsection*{Inflaton potential and backreaction}

Let us now analyse the effect of moduli stabilisation and backreaction during inflation. First notice that, even in this supergravity description, the kinetic term for the inflaton candidate $\phi = \re\, \Phi'$ depends on itself due to the breaking of the shift symmetry. The definition of the canonically normalised variable 
\be
\varphi = \int \sqrt{2 K_{\Phi \bar{\Phi}}}\, d\phi \, ,
\label{canorv}
\ee
is non-trivial. In particular, for the case at hand we see that 
\be
\sqrt{2 K_{\Phi \bar{\Phi}}} = \frac{\sqrt{su +\varepsilon (1+2 \varepsilon) \phi^2}}{su + \varepsilon \phi^2}\, .
\ee
which admits an analytic integral but it does not admit an analytic inverse. However, since $|\varepsilon| \ll 1$ we may approximate this expression by
\be
\sqrt{2 K_{\Phi \bar{\Phi}}}\, \simeq \, (su +\varepsilon \phi^2)^{-1/2}\left( 1 + \frac{\varepsilon^2\phi^2}{su + \varepsilon \phi^2}\right)   = \frac{1}{ \sqrt{s u}} \left( 1 -\frac{\varepsilon \phi^2}{2 s u} \right)  + \CO(\varepsilon^2)
\ee
where in the second equality we have expanded around $\varepsilon =0$. Integrating the last expression we arrive to
\be
\varphi = \frac{\phi}{\sqrt{su}} \left( 1 - \frac{\varepsilon \phi^2}{6su}\, \right) \, ,
\ee
whose inverse involves roots of a polynomial of degree 3. Since this effective 4d supergravity description is supposed to be valid in the small field limit we may assume that 
\be
|\varepsilon| \phi^2 \ll 6 s u \rightarrow \phi \sim \sqrt{su}\, \varphi \, ,
\label{normalization}
\ee
and use this relation in the following. 

\begin{figure}[H]
		\begin{center}
		\includegraphics[width=80mm]{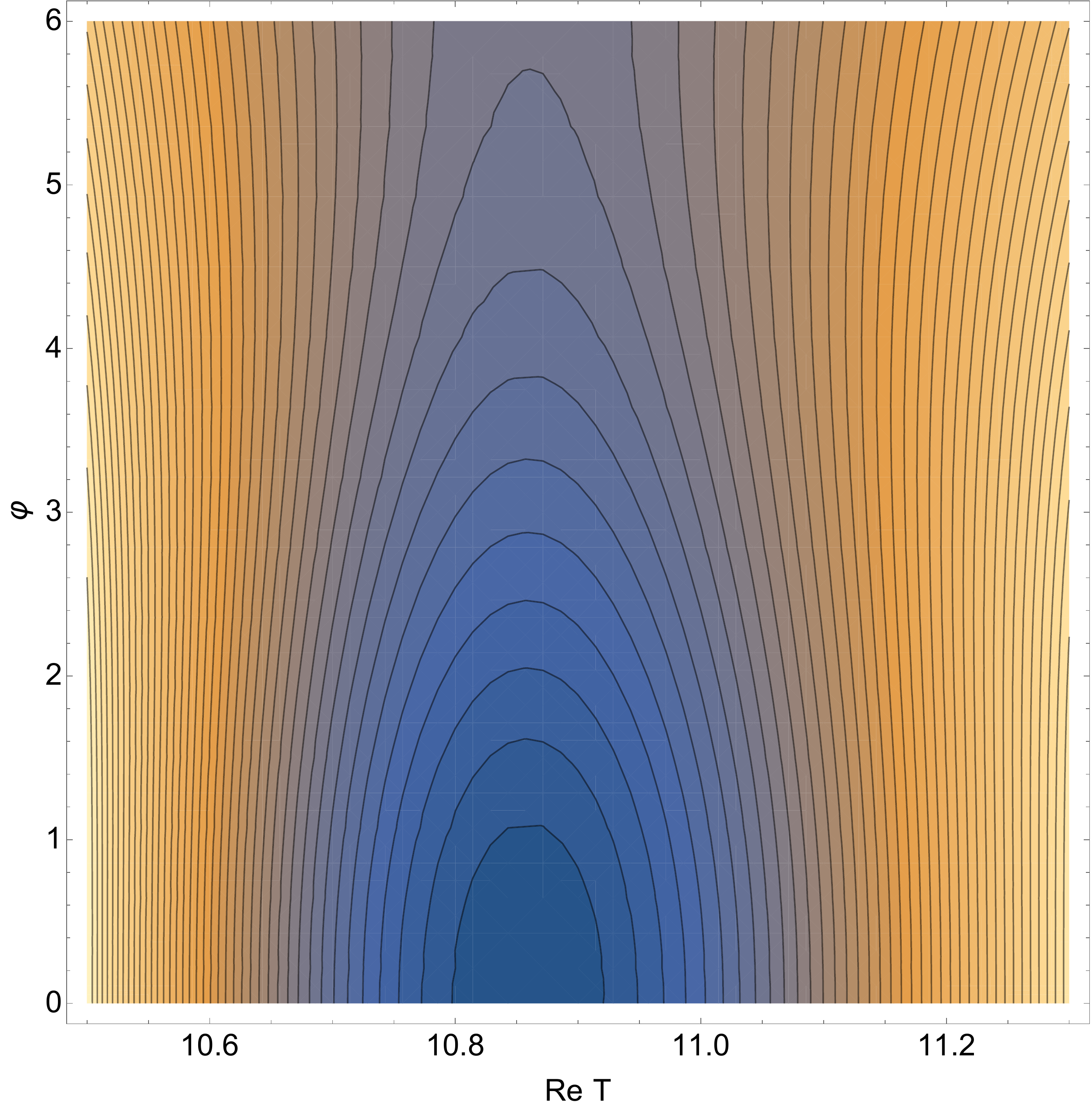} 
		\caption{Scalar potential evaluated in the $({\rm Re}\, T, \varphi)$ plane for the set of parameters (\ref{parameters}) where colder colours mean smaller values of $V$.}\label{fig5}
	\end{center}
	\end{figure}

Let us now address the backreaction effects of the K\"ahler modulus and the inflaton partner $\xi$. For this we will employ perturbation theory, where we define

\be
\re \, T  = t + \delta{\rm Re} T (\varphi) \text{  ,  } \im \, T  = 0 + \delta {\rm Im} T (\varphi) \text{  ,  } \xi = \langle \xi \rangle + \delta \xi(\varphi)\, ,
\ee
with $t$,  and $\langle \xi \rangle = 0$ are vevs of the backreacting fields in the Minkowski vacuum. Assuming that the fluctuations are small and minimising the scalar potential for them we find that
\be
\delta {\rm Re} T (\varphi) = \frac{3 \varepsilon ^2 \varphi^2}{2 a^3 t^2} + \mathcal{O}\left(\frac{H^2}{m_{T}^2}\right) \text{  ,  }  \delta  {\rm Im} T (\varphi) = 0  \text{  ,  } \delta \xi(\varphi) = 0 \, .
\ee
Notice that the backreaction of $\re\, T$ is suppressed by a factor of $t_{}^2$ as compared to similar setups, like e.g. in \cite{Bielleman:2016olv}. The main reason is that in our setup the K\"ahler modulus is not coupled to the inflaton neither via the superpotential nor the kinetic terms. It is only coupled via the overall factor of $e^{K}$ in the scalar potential. One way to check the consistency of this result is to plot the scalar potential in the plane $({\rm Re}\, T, \varphi)$ for the benchmark set of parameter values (\ref{parameters}), as done in figure \ref{fig5}. Indeed, there we see that the trajectory of minimum energy (represented by the darkest blue colour) is at this level of approximation a straight line in the $({\rm Re}\, T, \varphi)$ plane. This means that the K\"ahler modulus backreaction effects are essentially negligible. Numerically we have that 
\be
\delta {\rm Re} T (\varphi) \sim 10^{-4} \varphi^2\, ,
\label{backexample}
\ee
and the leading order contribution in the scalar potential will be $V_{back} \sim -1.55 \times 10^{-16} \varphi^4$.
%
%

The scalar potential taking into account both backreaction effects and the flattening induced by the kinetic term is then
\be
V \, =\, \frac{\varepsilon^2 W_{0}^2}{16 u s t^3}\left[\varphi^2  - 2 \varepsilon \varphi^4\right] +\,  \mathcal{O}\left(\varepsilon^4,\frac{1}{t^4}\right)\, ,
\label{scalarpot}
\ee
where the $\varphi^4$ term in the former expression arises only due to the non-trivial kinetic term, and not to the backreaction of heavy moduli. Unfortunately, when we plug the set of parameters (\ref{parameters}) into this potential we find a supergravity model where the slow-roll conditions cannot occur for more than $\Delta \varphi \sim 6 M_P$ and so the necessary number of e-folds cannot be attained. Of course, this supergravity description is only valid for the small-field limit. At large-field values we should not trust the supergravity scalar potential, which should be replaced by the DBI potential of section \ref{sec:flatflux}. By the analysis of subsection \ref{ssec:cosmob} we obtain that the corresponding flux-flattened potential would indeed attain the 60 e-fold of inflation with cosmological observables within current experimental bounds. The above analysis should then be understood as a means to estimate the magnitude of the backreaction effects. Indeed, if this magnitude is already negligible for (\ref{scalarpot}) we expect it to be even less important for the DBI scalar potential, since the effect of flux-flattening will lower the potential energy. We have found this to be a general feature of the effective supergravity models of the kind (\ref{KWMS}), irrespective of the set of effective parameters chosen. In fact, for a different choice of parameters one may easily construct models where 60 e-folds of inflation are attained and with realistic cosmological observables, already at the supergravity level.\footnote{Indeed, had we chosen the set of parameters 
\be\nonumber
\kappa_4 A = -1.05\,, \quad a = \frac{2 \pi}{26}\,, \quad\kappa_4 W_{0} = 0.48  \,, \quad su = 1.05 \,, \quad \varepsilon = 6.3 \times 10^{-4} \,, \quad t = 9.27 \,, \quad \Delta^2 = 0.28 \, ,
\ee
we would have also found mass scales similar to (\ref{benchmasses}) and a supergravity potential of the form (\ref{scalarpot}). However this potential would now be such that 60 e-folds are attained starting from $\varphi_{\star} = 14.16 M_{P}$, and with CMB observables with values $r = 0.069$ and $n_{s} = 0.960$. Again, the backreaction effects will be negligible, more precisely of the order $V_{back} \sim -3.13 \times 10^{-18} \varphi^4$. Hence, this example constitutes a 4d supergravity model of large-field inflation of interest on its own.}

\section{Conclusions}
\label{sec:conclu}

In this paper we have analysed an interesting class of models of F-term axion monodromy inflation \cite{Marchesano:2014mla} that arise in type IIB/F-theory flux compactifications with mobile D7-branes, of the kind proposed in \cite{Hebecker:2014eua,Ibanez:2014kia,Ibanez:2014swa}. The main observation that has triggered our analysis is that the flux-induced potential on the D7-brane position field $\Phi$ presents large flattening effects at large field values, due to the structure of the DBI+CS action. Flattening effect of this kind had already been observed in \cite{Ibanez:2014swa}, although for a very particular class of background fluxes compatible with the symmetries of the model. We have found that when one considers the most generic flux background, as one would need to do for the models in \cite{Hebecker:2014eua}, the flattening effects are much larger. This effect, dubbed {\em flux flattening}, arises due to the different dependence that the inflaton potential and kinetics terms have on $\Phi$ in the presence of generic background fluxes. It occurs that the kinetic terms grow equally or faster than the potential and so, upon canonical normalisation and at large field values, we find a potential that displays either a linear or smaller power-law behaviour. In fact, we have argued that in setups where $\Phi$ is lighter than the complex structure moduli of the compactification the latter case will apply, rendering flux flattening effects quite generic. Moreover, in simple setups where one of the component of the complex field $\Phi$ is much lighter than the other, we obtain a rather simple single-field inflationary model whose potential only depends on one parameter $\hat{\Upsilon}$. We have made a rough estimate for the range of values of this parameter and have shown that the related potentials nicely reproduce CMB observables within the current experimental bounds, attaining values for the tensor-to-scalar ratio as low as $r \sim 0.04$. 

In order to perform these estimates, and in order to determine how feasible are the scale hierarchies that lead to  single-field inflationary models, we have considered the embedding of mobile 7-branes with periodic directions in type IIB/F-theory flux compactifications. As in \cite{Hebecker:2014eua,Arends:2014qca} we have used the example of F-theory on ${\bf K3} \times {\bf K3}$ to develop our intuition on this system, and in particular on which kind of discrete and continuous symmetries will it exhibit. This picture has served to formulate under which conditions the 4d supergravity scalar potential of a compactification with a mobile D7-brane will contain a flat or a very light direction involving a particular component of $\Phi$, which we then identify with the inflaton field. In fact, we have found that the corresponding inflationary trajectory also involves large displacements of the inflaton field $S$. 
Armed with an approximate flat direction,  it is no longer necessary to tune any of the quantities of the superpotential (\ref{Weffp}) to a very small value like in \cite{Bielleman:2016olv} in order to obtain a low inflaton mass. Instead, we only need to require that the complex structure moduli functions $f$ and $g$ have a relatively similar phase when evaluated at the vacuum. We find that this milder requirement constitutes an advantage in order to build models of D7-brane chaotic inflation in general Calabi-Yau compactification, where $f$ and $g$ would depend on the period integrals. Finally, one may use this 4d supergravity description to combine the D7-brane system with the necessary ingredients to achieve K\"ahler moduli stabilisation. We have done so in a KKLT-like scheme with a single K\"ahler modulus, finding that K\"ahler moduli stabilisation is compatible with our single-field inflationary scenario for the hierarchy of scalar masses that the latter entails. 

There is a series of directions which would deserve further attention in order to render our flux flattening scenario more precise. First, as pointed out in \cite{Bielleman:2016olv}, including K\"ahler moduli stabilisation will induce the presence of imaginary anti-self-dual (IASD) background fluxes, which will in turn modify the DBI+CS D7-brane action. Since in our supergravity analysis the backreaction effects of K\"ahler moduli are negligible for our setup, we expect the same to be true for the contribution of IASD fluxes. Nevertheless, it would be interesting to generalise the D7-brane action computation of section \ref{sec:flatflux} to include the presence of IASD fluxes in order to directly verify this expectation. Moreover, in order to perform a more accurate analysis of backreaction effects along the inflationary trajectory, it would be interesting to describe the DBI+CS D7-brane potential and kinetic terms purely in terms of 4d supergravity, as done in \cite{Bielleman:2016grv,Bielleman:2016olv} for the Higgs-otic scenario. Due to the complicated square root dependence that arises due to the DBI action this seems in general quite a challenging task, but it may be achievable for the simplified expressions that arise for the choice of parameters made in subsection \ref{ssec:estimate}. Finally, we have analysed the compatibility with K\"ahler moduli stabilisation in a very particular KKLT-like scheme. Since in our models $W_0$ is not very small, one may also consider combining the D7-brane system with the moduli stabilisation scheme of the Large Volume Scenario \cite{Balasubramanian:2005zx}, as already suggested in \cite{Hebecker:2014eua}. 

One aspect that we have not considered is the backreaction of the inflaton field on the complex structure moduli of the compactification. As pointed out in \cite{Baume:2016psm,Klaewer:2016kiy,Valenzuela:2016yny,Blumenhagen:2017cxt}, such effects may severely reduce the canonical field distance and therefore prevent the inflaton field to perform the necessary excursion to attain 60 e-folds of inflation. It would be interesting to compute such backreaction effects, either by specifying the complex structure moduli functions in (\ref{Weffp}) in a explicit example and integrating them out together with $U$ and $S$, or by considering an intermediate effective theory with $S$, $U$ and $\Phi$ that contains the symmetries inherent to our scenario. In any event, we expect the general point made in \cite{Valenzuela:2016yny,Bielleman:2016olv} to also hold in our scenario. Namely, that these dangerous effects should be absent when achieving a appropriate hierarchy of masses between the inflaton and the closed-string moduli entering its kinetic term. Recently, it has been argued that creating such a hierarchy may lead to other problems like a flux scale above the Kaluza-Klein scale \cite{Blumenhagen:2017cxt}. In that respect, it is worth pointing out that in our setup there is  no need to make flux quanta large to achieve a low inflaton mass, and that the smallness of the quantities $\varepsilon$ and $W_0$ that create the mass hierarchy may easily arise from their dependence on complex structure moduli of the compactification on which the inflaton kinetic term does not depend. Therefore such considerations would a priori not apply to our scenario. At any rate, our proposed flux flattening effects and their embeddings into moduli stabilisation schemes that we presented provide an interesting arena for these backreaction issues to be concretely addressed. We hope to return to these issues in the future.


\bigskip

\centerline{\bf \large Acknowledgments}

\bigskip

We would like to thank L.E. Ib\'a\~nez, I.~Valenzuela, T.~Weigand and C.~Wieck for useful discussions. 
This work has been partially supported by the grants FPA2015-65480-P from  MINECO, SEV-2012-0249 of the ``Centro de Excelencia Severo Ochoa" Programme, the ERC Advanced Grant SPLE under contract ERC-2012-ADG-20120216-320421, the DOE grant DE-FG-02-95ER40896, the Kellett Award of the University of Wisconsin, and the HKRGC grants HUKST4/CRF/13G and 16304414.
A.L. is supported through the FPI grant SVP-2013-067949 and EEBB-I-2016-11573 from MINECO. A.L., F.M. and G.Z. would like to thank UW-Madison for hospitality during completion of this work. G.Z. would also like to thank ITF at Utrecht University and Padova University for hospitality during completion
of this work.


\appendix

\section{$SL(2,\IR)$ transformations of the K\"ahler and superpotential and alternative effective theories}
\label{ap:SL}

Let us consider a K\"ahler potential of the form
\be
K \, =\,  - \log\left[(\Phi-\overline \Phi)^2-(S-\overline S)(U-\overline U)\right] + K_2
\label{ap:Keffp}
\ee
where $K_2$ does not contain any dependence on $U,S,\Phi$. Then following \cite{LopesCardoso:1994is,Antoniadis:1994hg,Brignole:1995fb,Brignole:1996xb} we see that $K$ is invariant under a $SL(2,\IR)_U$ symmetry up to a K\"ahler transformation. More precisely we have that by under the following field redefinitions
\bea
\label{ULt}
U & \rightarrow & \frac{aU + b}{cU + d} \\
\label{SLt}
S & \rightarrow & S - c\frac{\Phi^2}{cU + d} \\
\Phi & \rightarrow & \frac{\Phi}{cU+d}
\label{PhiLt}
\eea
with $a,b,c,d \in \IR$ and $ad-bc = 1$, the K\"ahler potential transforms as
\be
K \rightarrow K + \log|d+cU|^2\, .
\label{ap:Ktrans}
\ee
Let us now take a superpotential of the form 
\be
W = \hat{\mathfrak{n}} + \hat{\mathfrak{m}} \, U -{\mathfrak{n}}\, S + {\mathfrak{m}} \left(\Phi^2-S U\right)+ 2\mathfrak{f}\, \Phi   + W_2
\label{ap:Weff}
\ee
where $W_2$ and the calligraphic letters are functions of other moduli but not of $U,S,\Phi$. Applying the above set of field redefinitions and taking into account the K\"ahler transformation (\ref{ap:Ktrans}) we obtain
\be
W \rightarrow W' \, =\, \hat{\mathfrak{n}}' + \hat{\mathfrak{m}}' \, U -{\mathfrak{n}}'\, S + {\mathfrak{m}}' \left(\Phi^2-S U\right)+ 2\mathfrak{f}\, \Phi  + (cU+d) W_2
\ee
where
\be
{\mathfrak{n}}' = d\mathfrak{n} + b\mathfrak{m}\, , \quad   {\mathfrak{m}}' = a \mathfrak{m}+c \mathfrak{n} \, , \quad \hat{\mathfrak{n}}' = d\hat{\mathfrak{n}} + b\hat{\mathfrak{m}}\, , \quad \hat{\mathfrak{m}}' = a\hat{\mathfrak{m}} +  c\hat{\mathfrak{n}}\, .
\ee
In particular, if $\mathfrak{n}$ and $\mathfrak{m}$ have the same phase we can always choose $a$ and $c$ such that ${\mathfrak{m}}' = 0$. 
In this case, for $\mathfrak{f} =0$  we have a flat direction along $\re\, \Phi$. One can then see that, in terms of the original variables this precisely corresponds to the trajectory (\ref{Kshift}), with $r/s = -c/a$.

Interestingly, one can use this freedom to obtain an expression for $W$ and $K$ more suitable for the purposes of section \ref{ssec:moduli}, namely to obtain an effective theory for the fields $\Phi$ and $T$ in order to analyse moduli stabilisation. For this, recall that $\mathfrak{n}$, $\mathfrak{m}$, $\hat{\mathfrak{n}}$, $\hat{\mathfrak{m}}$ are functions of the complex structure moduli of the compactification. Let us now denote their numerical value at the vacuum $\Phi=0$ by their non-calligraphic version. That is,
\be
n = \mathfrak{n}|_{\Phi =0}\, , \qquad m = \mathfrak{m}|_{\Phi =0}\, , \qquad \hat{n} = \hat{\mathfrak{n}}|_{\Phi =0}\, ,  \qquad \hat{m} = \hat{\mathfrak{m}}|_{\Phi =0} \, .
\ee
Now, as these quantities are numbers we can do the field redefinition (\ref{ULt}-\ref{PhiLt}) with parameters
\be
a\, =\, 1 \qquad b\, =\, 0 \qquad c\, =\, - \re\, \left( \frac{m}{n} \right) \qquad d \, = \, 1\, .
\ee
In terms of the new variables we have the same K\"ahler potential (\ref{ap:Keffp}), and the new superpotential 
\be
W \, =\, \hat{\mathfrak{n}} + \left( \hat{\mathfrak{m}} - \re \left( \frac{m}{n} \right)  \hat{\mathfrak{n}} \right) \, U -{\mathfrak{n}} \, S + \left({\mathfrak{m}} - \re \left( \frac{m}{n} \right) {\mathfrak{n}}\right) \left(\Phi^2-S U\right)+ \dots
\ee
and so, if we write this superpotential in the form (\ref{Weffp}) we have that at the vacuum
\be
\frac{g}{f} \, =\, i \im \,  \left( \frac{m}{n} \right)\, =\, i \left| \frac{m}{n}\right| {\rm sin \, } \left( \th_n - \th_m\right)
\ee
where $\th_m$, $\th_n$ are the phases of $m$ and $n$, respectively. By our assumptions of the main text this difference of phases is very small and so this is a very small number. We then recover a shift-symmetric K\"ahler potential and a superpotential with new modulus dependent coefficients. Near the vacuum the coefficient for $\Phi^2$ is much smaller than those for the closed string moduli, and a slight misalignment of phases plays the role of an effective $\mu$-term. This $\mu$-term is in particular much smaller than the coefficient of $S$ and with a phase that differs by $e^{i\pi/2}$. Under these circumstances it seems quite reasonable to apply the strategy of \cite{Landete:2016cix} to the new complex structure and dilaton $S$, with the latter differing slightly from the variable (\ref{hatS}). After that we obtain an effective theory for $\Phi$ given by
\be
W\, =\, W_0 + \mu \Phi^2 +\dots \qquad \qquad \mu = i n \im \,  \left( \frac{m}{n} \right)
\ee
and a K\"ahler potential of the form (\ref{ap:Keffp}), where now $S$ and $U$ are replaced by their vevs. As in section \ref{ssec:moduli} one may add the contribution from the K\"ahler moduli sector to address full moduli stabilisation below the flux scale. For instance, in a KKLT-like scenario one would obtain an effective potential of the form
\be
W\, =\, W_0 + \mu \Phi^2 + Ae^{-aT}
\ee
and a K\"ahler potential given by
\be
K \, =\,  - 3 \log\left[T+\bar{T}\right] - \log\left[4su + (\Phi-\overline \Phi)^2\right] 
\ee
with $s = \langle \im\, S \rangle$ and $u = \langle \im\, U \rangle$. The computational details of the complex structure and K\"ahler moduli backreaction and the conditions needed in order to have trans-Planckian field ranges are then similar to the ones discussed in \cite{Bielleman:2016olv}.

\section{Other single field potentials}
\label{ap:other}

\begin{center}
\begin{figure}[H]
\includegraphics[width=80mm]{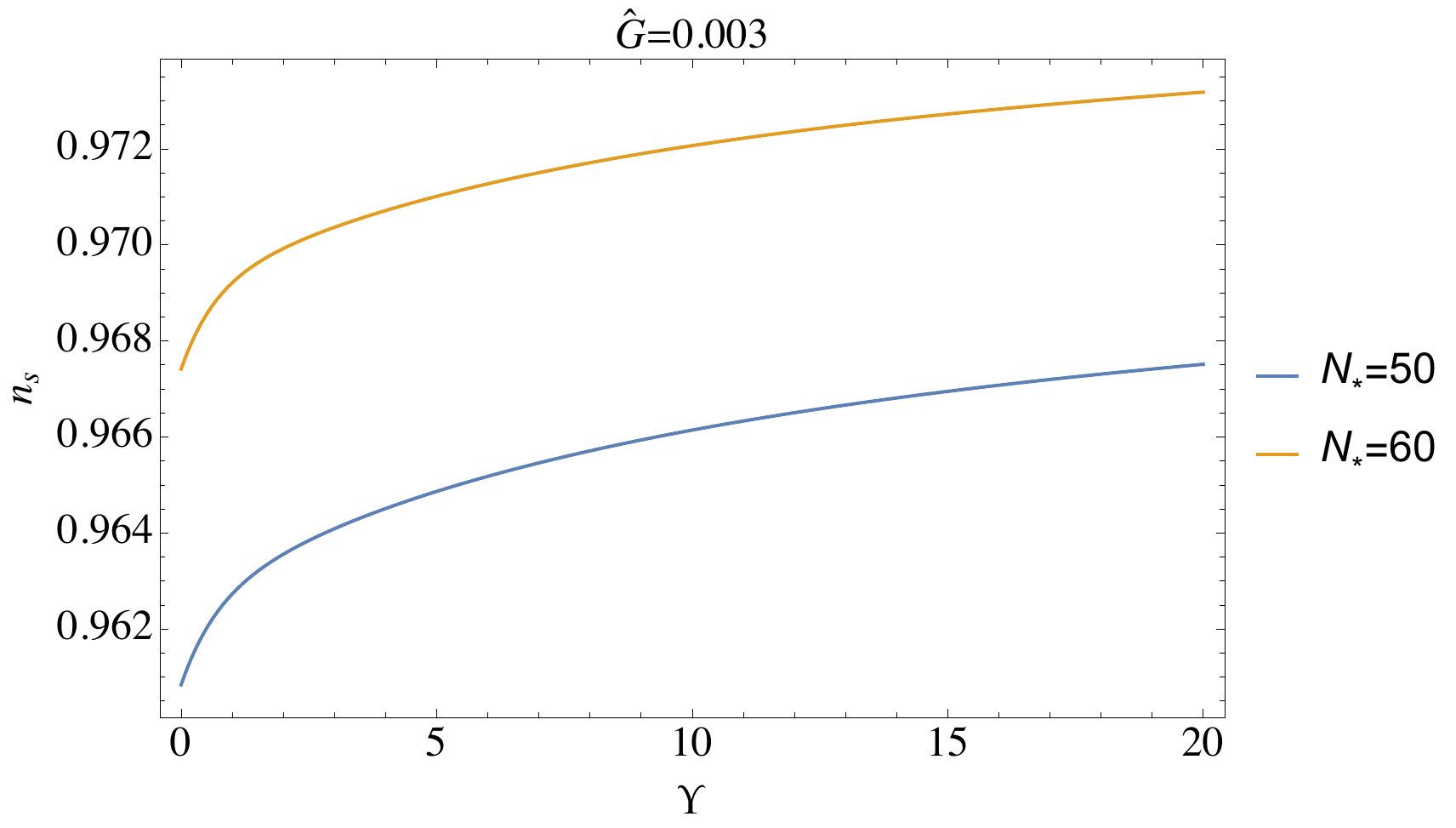} \includegraphics[width=80mm]{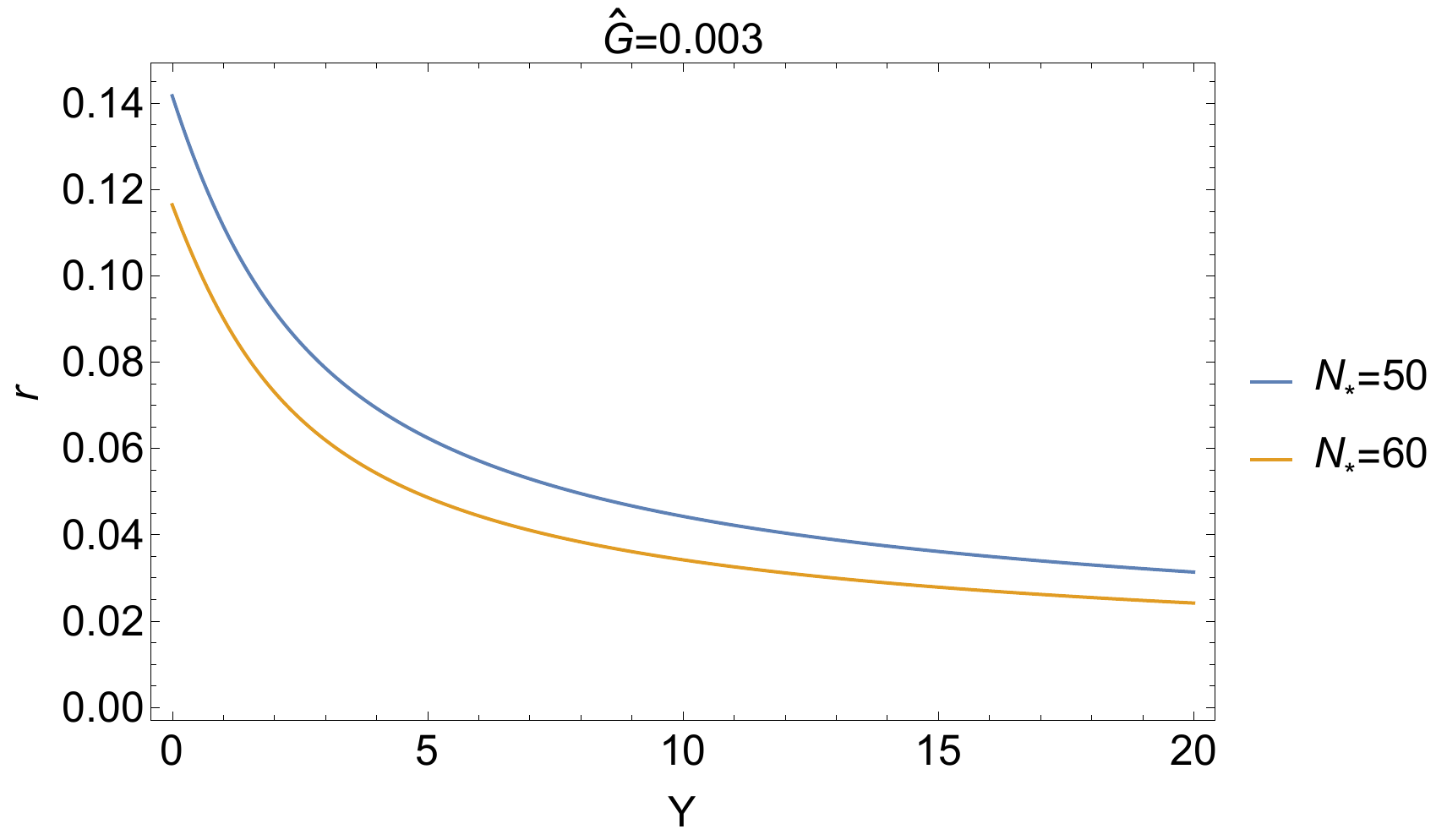}
\vspace{-.75cm}
\caption{Spectral index $n_s$ and tensor-to-scalar ratio $r$ in terms of $\Upsilon$ with $\hat G =0.003$ for $N_*=50$ and $N_* = 60$ e-folds.}\label{fig2}
\end{figure}
\end{center}

\begin{center}
\begin{figure}[H]
\includegraphics[width=80mm]{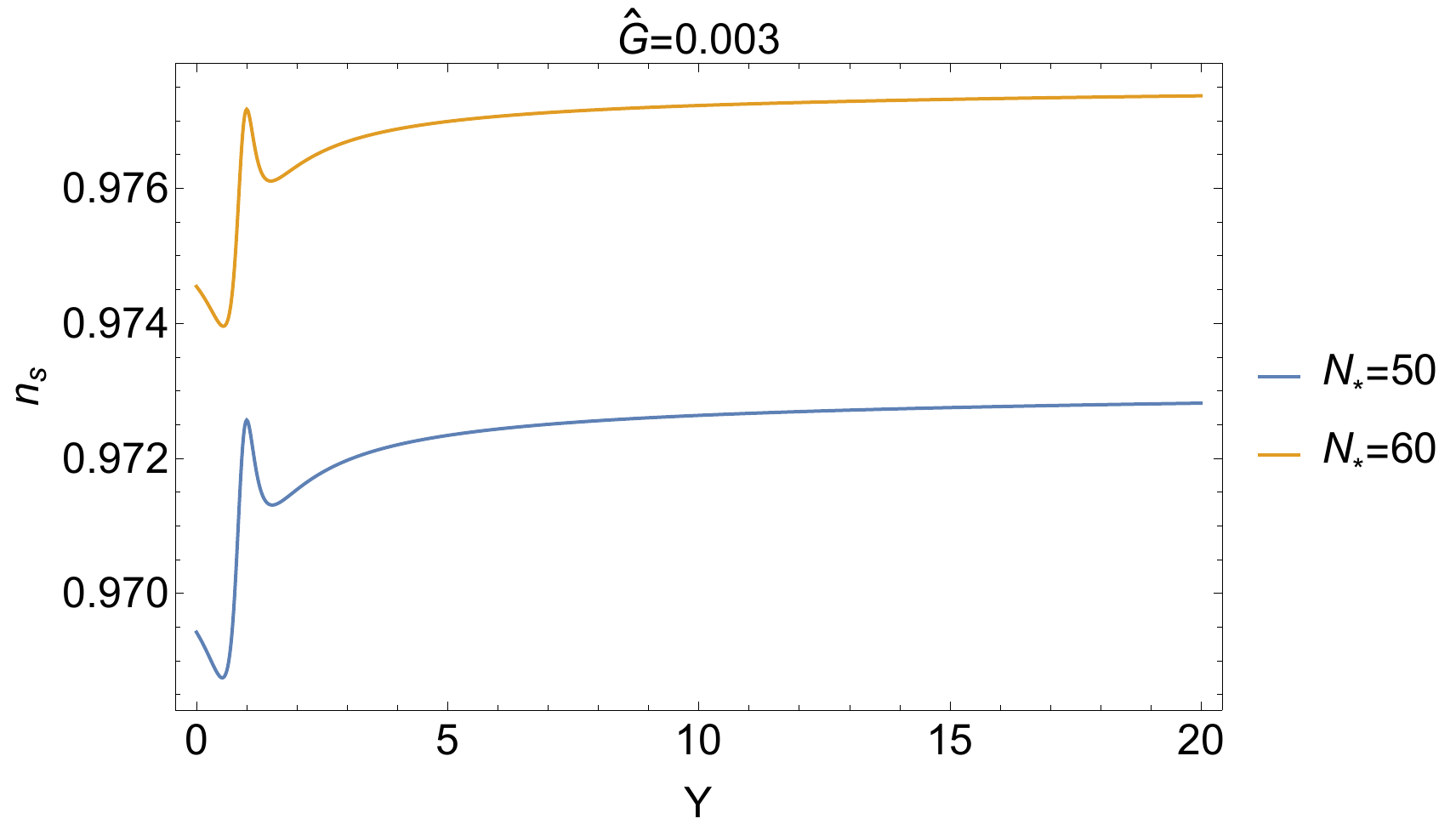} \includegraphics[width=80mm]{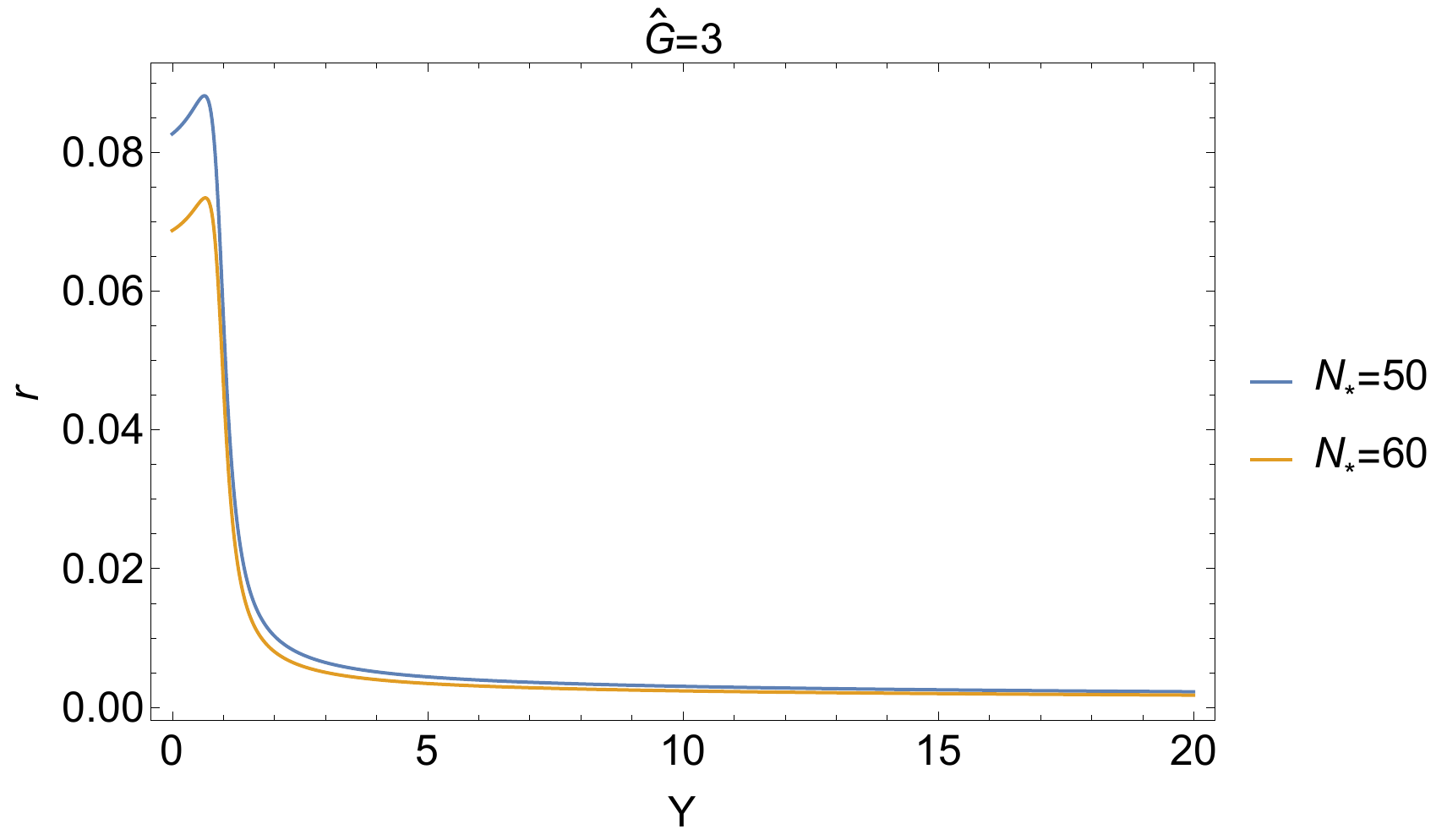}
\vspace{-.75cm}
\caption{$n_s$ and $r$ in terms of $\Upsilon$ with $\hat G =3$ for $N_*=50$ and $N_* = 60$ e-folds.}\label{fig3}
\end{figure}
\end{center}

\begin{figure}[H]
	\begin{center}
\includegraphics[width=130mm]{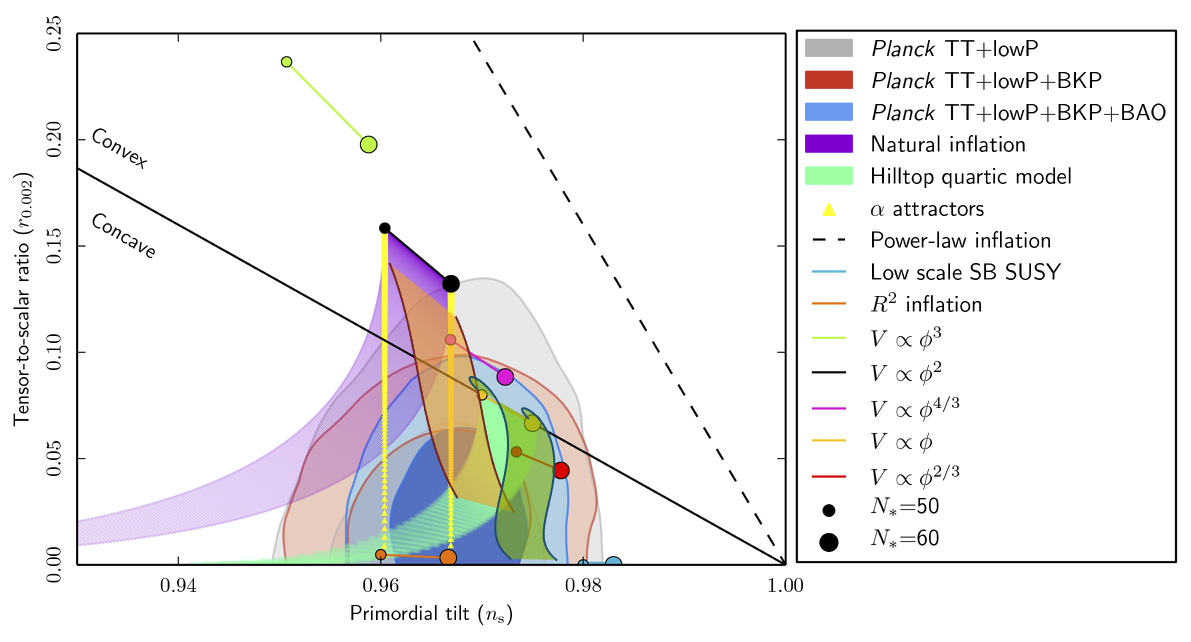}
\vspace{-.5cm}
\caption{Region for the spectral index $n_s$ vs tensor-to-scalar ratio $r$ 
for the two values of $\hat G$ (orange region corresponds to $\hat G = 0.003$ and green to $\hat G = 3$) and $0\leqslant \Upsilon \leqslant 20$.}\label{fig4}
\end{center}
\end{figure}

In this appendix we perform an analysis for the D7-brane single field potential of subsection \ref{ssec:asympto} along the lines of subsection \ref{ssec:cosmob}, but for different values of $\hat G$ and $\Upsilon$ that may arise in different setups from the one of subsection \ref{ssec:moduli}. We considered two regions in the $\hat G$ parameter space, namely $\hat G \sim 0.003$ and $\hat G \sim 3$, and vary $\Upsilon$ which is the parameter that controls the deviation from the model of \cite{Ibanez:2014swa}. We show how the cosmological observables vary in the two regimes for  $0 \leqslant \Upsilon \leqslant 20$ in the figures \ref{fig2} and \ref{fig3} .
%
%
\vspace{-.5cm}
%
%
\vspace{-.75cm}
We see that in both cases the effect of the parameter $\Upsilon$ is quite dramatic: it leads to a significant lowering of the tensor-to-scalar ratio $r$ as expected from the flattening 
induced by the self-dual component of the flux $\mathcal F$. At the same time the spectral index $n_s$ generally moves closer to 1 as $\Upsilon$ increases. This behaviour occurs
for both regimes of $\hat G$ that we chose to explore. The r\^ole of this second parameter is to provide (at $\Upsilon = 0$) an interpolation between models with quadratic
and linear potential as already observed in\cite{Ibanez:2014swa} (a similar interpolation between quadratic and linear potentials was also observed in \cite{Escobar:2015fda,Escobar:2015ckf}). Therefore if we allow for more general values of $\hat G$ and $\Upsilon$ than the ones used in section \ref{ssec:estimate} we see that it is possible to explore additional
regions of the $n_s -  r$ plane, namely we may start with any potential interpolating between quadratic and linear (the exact interpolation being set roughly by $\hat G$) and by increasing $\Upsilon$ access regions with a lower value of the tensor-to-scalar ratio $r$.
To show this more explicitly 
we chose to superimpose over the Planck collaboration results \cite{Ade:2015lrj} the two regions explored in the $n_s - r$ plane, showing the result in figure \ref{fig4}.
%
%

\bibliographystyle{JHEP}
\bibliography{refs}

\providecommand{\href}[2]{#2}\begingroup\raggedright\begin{thebibliography}{10}

\bibitem{Array:2015xqh}
{\bf BICEP2, Keck Array} Collaboration, P.~A.~R. Ade {\em et.~al.}, {\it
  {Improved Constraints on Cosmology and Foregrounds from BICEP2 and Keck Array
  Cosmic Microwave Background Data with Inclusion of 95 GHz Band}},  {\em Phys.
  Rev. Lett.} {\bf 116} (2016) 031302,
  [\href{http://xxx.lanl.gov/abs/1510.09217}{{\tt 1510.09217}}].

\bibitem{Lyth:1996im}
D.~H. Lyth, {\it {What would we learn by detecting a gravitational wave signal
  in the cosmic microwave background anisotropy?}},  {\em Phys. Rev. Lett.}
  {\bf 78} (1997) 1861--1863,
  [\href{http://xxx.lanl.gov/abs/hep-ph/9606387}{{\tt hep-ph/9606387}}].

\bibitem{ArkaniHamed:2006dz}
N.~Arkani-Hamed, L.~Motl, A.~Nicolis, and C.~Vafa, {\it {The String landscape,
  black holes and gravity as the weakest force}},  {\em JHEP} {\bf 06} (2007)
  060, [\href{http://xxx.lanl.gov/abs/hep-th/0601001}{{\tt hep-th/0601001}}].

\bibitem{delaFuente:2014aca}
A.~de~la Fuente, P.~Saraswat, and R.~Sundrum, {\it {Natural Inflation and
  Quantum Gravity}},  {\em Phys. Rev. Lett.} {\bf 114} (2015), no.~15 151303,
  [\href{http://xxx.lanl.gov/abs/1412.3457}{{\tt 1412.3457}}].

\bibitem{Rudelius:2015xta}
T.~Rudelius, {\it {Constraints on Axion Inflation from the Weak Gravity
  Conjecture}},  {\em JCAP} {\bf 1509} (2015), no.~09 020,
  [\href{http://xxx.lanl.gov/abs/1503.00795}{{\tt 1503.00795}}].

\bibitem{Montero:2015ofa}
M.~Montero, A.~M. Uranga, and I.~Valenzuela, {\it {Transplanckian axions!?}},
  {\em JHEP} {\bf 08} (2015) 032,
  [\href{http://xxx.lanl.gov/abs/1503.03886}{{\tt 1503.03886}}].

\bibitem{Brown:2015iha}
J.~Brown, W.~Cottrell, G.~Shiu, and P.~Soler, {\it {Fencing in the Swampland:
  Quantum Gravity Constraints on Large Field Inflation}},  {\em JHEP} {\bf 10}
  (2015) 023, [\href{http://xxx.lanl.gov/abs/1503.04783}{{\tt 1503.04783}}].

\bibitem{Brown:2015lia}
J.~Brown, W.~Cottrell, G.~Shiu, and P.~Soler, {\it {On Axionic Field Ranges,
  Loopholes and the Weak Gravity Conjecture}},  {\em JHEP} {\bf 04} (2016) 017,
  [\href{http://xxx.lanl.gov/abs/1504.00659}{{\tt 1504.00659}}].

\bibitem{Hebecker:2015rya}
A.~Hebecker, P.~Mangat, F.~Rompineve, and L.~T. Witkowski, {\it {Winding out of
  the Swamp: Evading the Weak Gravity Conjecture with F-term Winding
  Inflation?}},  {\em Phys. Lett.} {\bf B748} (2015) 455--462,
  [\href{http://xxx.lanl.gov/abs/1503.07912}{{\tt 1503.07912}}].

\bibitem{Bachlechner:2015qja}
T.~C. Bachlechner, C.~Long, and L.~McAllister, {\it {Planckian Axions and the
  Weak Gravity Conjecture}},  {\em JHEP} {\bf 01} (2016) 091,
  [\href{http://xxx.lanl.gov/abs/1503.07853}{{\tt 1503.07853}}].

\bibitem{Junghans:2015hba}
D.~Junghans, {\it {Large-Field Inflation with Multiple Axions and the Weak
  Gravity Conjecture}},  {\em JHEP} {\bf 02} (2016) 128,
  [\href{http://xxx.lanl.gov/abs/1504.03566}{{\tt 1504.03566}}].

\bibitem{Heidenreich:2015wga}
B.~Heidenreich, M.~Reece, and T.~Rudelius, {\it {Weak Gravity Strongly
  Constrains Large-Field Axion Inflation}},  {\em JHEP} {\bf 12} (2015) 108,
  [\href{http://xxx.lanl.gov/abs/1506.03447}{{\tt 1506.03447}}].

\bibitem{Silverstein:2008sg}
E.~Silverstein and A.~Westphal, {\it {Monodromy in the CMB: Gravity Waves and
  String Inflation}},  {\em Phys. Rev.} {\bf D78} (2008) 106003,
  [\href{http://xxx.lanl.gov/abs/0803.3085}{{\tt 0803.3085}}].

\bibitem{McAllister:2008hb}
L.~McAllister, E.~Silverstein, and A.~Westphal, {\it {Gravity Waves and Linear
  Inflation from Axion Monodromy}},  {\em Phys. Rev.} {\bf D82} (2010) 046003,
  [\href{http://xxx.lanl.gov/abs/0808.0706}{{\tt 0808.0706}}].

\bibitem{Marchesano:2014mla}
F.~Marchesano, G.~Shiu, and A.~M. Uranga, {\it {F-term Axion Monodromy
  Inflation}},  {\em JHEP} {\bf 09} (2014) 184,
  [\href{http://xxx.lanl.gov/abs/1404.3040}{{\tt 1404.3040}}].

\bibitem{Blumenhagen:2014gta}
R.~Blumenhagen and E.~Plauschinn, {\it {Towards Universal Axion Inflation and
  Reheating in String Theory}},  {\em Phys. Lett.} {\bf B736} (2014) 482--487,
  [\href{http://xxx.lanl.gov/abs/1404.3542}{{\tt 1404.3542}}].

\bibitem{Hebecker:2014eua}
A.~Hebecker, S.~C. Kraus, and L.~T. Witkowski, {\it {D7-Brane Chaotic
  Inflation}},  {\em Phys. Lett.} {\bf B737} (2014) 16--22,
  [\href{http://xxx.lanl.gov/abs/1404.3711}{{\tt 1404.3711}}].

\bibitem{Ibanez:2014kia}
L.~E. Ib\'a\~nez and I.~Valenzuela, {\it {The inflaton as an MSSM Higgs and
  open string modulus monodromy inflation}},  {\em Phys. Lett.} {\bf B736}
  (2014) 226--230, [\href{http://xxx.lanl.gov/abs/1404.5235}{{\tt 1404.5235}}].

\bibitem{Franco:2014hsa}
S.~Franco, D.~Galloni, A.~Retolaza, and A.~Uranga, {\it {On axion monodromy
  inflation in warped throats}},  {\em JHEP} {\bf 02} (2015) 086,
  [\href{http://xxx.lanl.gov/abs/1405.7044}{{\tt 1405.7044}}].

\bibitem{Blumenhagen:2014nba}
R.~Blumenhagen, D.~Herschmann, and E.~Plauschinn, {\it {The Challenge of
  Realizing F-term Axion Monodromy Inflation in String Theory}},  {\em JHEP}
  {\bf 01} (2015) 007, [\href{http://xxx.lanl.gov/abs/1409.7075}{{\tt
  1409.7075}}].

\bibitem{Hayashi:2014aua}
H.~Hayashi, R.~Matsuda, and T.~Watari, {\it {Issues in Complex Structure Moduli
  Inflation}},  \href{http://xxx.lanl.gov/abs/1410.7522}{{\tt 1410.7522}}.

\bibitem{Hebecker:2014kva}
A.~Hebecker, P.~Mangat, F.~Rompineve, and L.~T. Witkowski, {\it {Tuning and
  Backreaction in F-term Axion Monodromy Inflation}},  {\em Nucl. Phys.} {\bf
  B894} (2015) 456--495, [\href{http://xxx.lanl.gov/abs/1411.2032}{{\tt
  1411.2032}}].

\bibitem{Ibanez:2014swa}
L.~E. Ib\'a\~nez, F.~Marchesano, and I.~Valenzuela, {\it {Higgs-otic Inflation
  and String Theory}},  {\em JHEP} {\bf 01} (2015) 128,
  [\href{http://xxx.lanl.gov/abs/1411.5380}{{\tt 1411.5380}}].

\bibitem{Garcia-Etxebarria:2014wla}
I.~Garc\'{\i}a-Etxebarria, T.~W. Grimm, and I.~Valenzuela, {\it {Special Points
  of Inflation in Flux Compactifications}},  {\em Nucl. Phys.} {\bf B899}
  (2015) 414--443, [\href{http://xxx.lanl.gov/abs/1412.5537}{{\tt 1412.5537}}].

\bibitem{Blumenhagen:2015kja}
R.~Blumenhagen, A.~Font, M.~Fuchs, D.~Herschmann, E.~Plauschinn, Y.~Sekiguchi,
  and F.~Wolf, {\it {A Flux-Scaling Scenario for High-Scale Moduli
  Stabilization in String Theory}},  {\em Nucl. Phys.} {\bf B897} (2015)
  500--554, [\href{http://xxx.lanl.gov/abs/1503.07634}{{\tt 1503.07634}}].

\bibitem{Escobar:2015fda}
D.~Escobar, A.~Landete, F.~Marchesano, and D.~Regalado, {\it {Large field
  inflation from D-branes}},  {\em Phys. Rev.} {\bf D93} (2016), no.~8 081301,
  [\href{http://xxx.lanl.gov/abs/1505.07871}{{\tt 1505.07871}}].

\bibitem{Escobar:2015ckf}
D.~Escobar, A.~Landete, F.~Marchesano, and D.~Regalado, {\it {D6-branes and
  axion monodromy inflation}},  {\em JHEP} {\bf 03} (2016) 113,
  [\href{http://xxx.lanl.gov/abs/1511.08820}{{\tt 1511.08820}}].

\bibitem{Hebecker:2015tzo}
A.~Hebecker, J.~Moritz, A.~Westphal, and L.~T. Witkowski, {\it {Towards Axion
  Monodromy Inflation with Warped KK-Modes}},  {\em Phys. Lett.} {\bf B754}
  (2016) 328--334, [\href{http://xxx.lanl.gov/abs/1512.04463}{{\tt
  1512.04463}}].

\bibitem{Bizet:2016paj}
N.~Cabo~Bizet, O.~Loaiza-Brito, and I.~Zavala, {\it {Mirror quintic vacua:
  hierarchies and inflation}},  {\em JHEP} {\bf 10} (2016) 082,
  [\href{http://xxx.lanl.gov/abs/1605.03974}{{\tt 1605.03974}}].

\bibitem{Landete:2016cix}
A.~Landete, F.~Marchesano, and C.~Wieck, {\it {Challenges for D-brane
  large-field inflation with stabilizer fields}},  {\em JHEP} {\bf 09} (2016)
  119, [\href{http://xxx.lanl.gov/abs/1607.01680}{{\tt 1607.01680}}].

\bibitem{Ibanez:2015fcv}
L.~E. Ib\'a\~nez, M.~Montero, A.~Uranga, and I.~Valenzuela, {\it {Relaxion
  Monodromy and the Weak Gravity Conjecture}},  {\em JHEP} {\bf 04} (2016) 020,
  [\href{http://xxx.lanl.gov/abs/1512.00025}{{\tt 1512.00025}}].

\bibitem{Hebecker:2015zss}
A.~Hebecker, F.~Rompineve, and A.~Westphal, {\it {Axion Monodromy and the Weak
  Gravity Conjecture}},  {\em JHEP} {\bf 04} (2016) 157,
  [\href{http://xxx.lanl.gov/abs/1512.03768}{{\tt 1512.03768}}].

\bibitem{Brown:2016nqt}
J.~Brown, W.~Cottrell, G.~Shiu, and P.~Soler, {\it {Tunneling in Axion
  Monodromy}},  {\em JHEP} {\bf 10} (2016) 025,
  [\href{http://xxx.lanl.gov/abs/1607.00037}{{\tt 1607.00037}}].

\bibitem{Dong:2010in}
X.~Dong, B.~Horn, E.~Silverstein, and A.~Westphal, {\it {Simple exercises to
  flatten your potential}},  {\em Phys. Rev.} {\bf D84} (2011) 026011,
  [\href{http://xxx.lanl.gov/abs/1011.4521}{{\tt 1011.4521}}].

\bibitem{Baumann:2014nda}
D.~Baumann and L.~McAllister, {\em {Inflation and String Theory}}.
\newblock Cambridge University Press, 2015.

\bibitem{Westphal:2014ana}
A.~Westphal, {\it {String cosmology Ñ Large-field inflation in string theory}},
   {\em Int. J. Mod. Phys.} {\bf A30} (2015), no.~09 1530024,
  [\href{http://xxx.lanl.gov/abs/1409.5350}{{\tt 1409.5350}}].

\bibitem{Silverstein:2016ggb}
E.~Silverstein, {\it {TASI lectures on cosmological observables and string
  theory}},  in {\em {Proceedings, Theoretical Advanced Study Institute in
  Elementary Particle Physics: New Frontiers in Fields and Strings (TASI 2015):
  Boulder, CO, USA, June 1-26, 2015}}, pp.~545--606, 2017.
\newblock \href{http://xxx.lanl.gov/abs/1606.03640}{{\tt 1606.03640}}.

\bibitem{Berg:2009tg}
M.~Berg, E.~Pajer, and S.~Sjors, {\it {Dante's Inferno}},  {\em Phys. Rev.}
  {\bf D81} (2010) 103535, [\href{http://xxx.lanl.gov/abs/0912.1341}{{\tt
  0912.1341}}].

\bibitem{Gur-Ari:2013sba}
G.~Gur-Ari, {\it {Brane Inflation and Moduli Stabilization on Twisted Tori}},
  {\em JHEP} {\bf 01} (2014) 179,
  [\href{http://xxx.lanl.gov/abs/1310.6787}{{\tt 1310.6787}}].

\bibitem{Palti:2014kza}
E.~Palti and T.~Weigand, {\it {Towards large r from [p, q]-inflation}},  {\em
  JHEP} {\bf 04} (2014) 155, [\href{http://xxx.lanl.gov/abs/1403.7507}{{\tt
  1403.7507}}].

\bibitem{Camara:2004jj}
P.~G. Camara, L.~E. Ibanez, and A.~M. Uranga, {\it {Flux-induced SUSY-breaking
  soft terms on D7-D3 brane systems}},  {\em Nucl. Phys.} {\bf B708} (2005)
  268--316, [\href{http://xxx.lanl.gov/abs/hep-th/0408036}{{\tt
  hep-th/0408036}}].

\bibitem{Gomis:2005wc}
J.~Gomis, F.~Marchesano, and D.~Mateos, {\it {An Open string landscape}},  {\em
  JHEP} {\bf 11} (2005) 021,
  [\href{http://xxx.lanl.gov/abs/hep-th/0506179}{{\tt hep-th/0506179}}].

\bibitem{Bielleman:2016olv}
S.~Bielleman, L.~E. Ib\'a\~nez, F.~G. Pedro, I.~Valenzuela, and C.~Wieck, {\it
  {Higgs-otic Inflation and Moduli Stabilization}},
  \href{http://xxx.lanl.gov/abs/1611.07084}{{\tt 1611.07084}}.

\bibitem{Giddings:2001yu}
S.~B. Giddings, S.~Kachru, and J.~Polchinski, {\it {Hierarchies from fluxes in
  string compactifications}},  {\em Phys. Rev.} {\bf D66} (2002) 106006,
  [\href{http://xxx.lanl.gov/abs/hep-th/0105097}{{\tt hep-th/0105097}}].

\bibitem{Ade:2015lrj}
{\bf Planck} Collaboration, P.~A.~R. Ade {\em et.~al.}, {\it {Planck 2015
  results. XX. Constraints on inflation}},  {\em Astron. Astrophys.} {\bf 594}
  (2016) A20, [\href{http://xxx.lanl.gov/abs/1502.02114}{{\tt 1502.02114}}].

\bibitem{Arends:2014qca}
M.~Arends, A.~Hebecker, K.~Heimpel, S.~C. Kraus, D.~Lust, C.~Mayrhofer,
  C.~Schick, and T.~Weigand, {\it {D7-Brane Moduli Space in Axion Monodromy and
  Fluxbrane Inflation}},  {\em Fortsch. Phys.} {\bf 62} (2014) 647--702,
  [\href{http://xxx.lanl.gov/abs/1405.0283}{{\tt 1405.0283}}].

\bibitem{Dasgupta:1999ss}
K.~Dasgupta, G.~Rajesh, and S.~Sethi, {\it {M theory, orientifolds and G -
  flux}},  {\em JHEP} {\bf 08} (1999) 023,
  [\href{http://xxx.lanl.gov/abs/hep-th/9908088}{{\tt hep-th/9908088}}].

\bibitem{Gorlich:2004qm}
L.~Gorlich, S.~Kachru, P.~K. Tripathy, and S.~P. Trivedi, {\it {Gaugino
  condensation and nonperturbative superpotentials in flux compactifications}},
   {\em JHEP} {\bf 12} (2004) 074,
  [\href{http://xxx.lanl.gov/abs/hep-th/0407130}{{\tt hep-th/0407130}}].

\bibitem{Lust:2005bd}
D.~Lust, P.~Mayr, S.~Reffert, and S.~Stieberger, {\it {F-theory flux,
  destabilization of orientifolds and soft terms on D7-branes}},  {\em Nucl.
  Phys.} {\bf B732} (2006) 243--290,
  [\href{http://xxx.lanl.gov/abs/hep-th/0501139}{{\tt hep-th/0501139}}].

\bibitem{Braun:2008ua}
A.~P. Braun, A.~Hebecker, and H.~Triendl, {\it {D7-Brane Motion from M-Theory
  Cycles and Obstructions in the Weak Coupling Limit}},  {\em Nucl. Phys.} {\bf
  B800} (2008) 298--329, [\href{http://xxx.lanl.gov/abs/0801.2163}{{\tt
  0801.2163}}].

\bibitem{Braun:2008pz}
A.~P. Braun, A.~Hebecker, C.~Ludeling, and R.~Valandro, {\it {Fixing D7 Brane
  Positions by F-Theory Fluxes}},  {\em Nucl. Phys.} {\bf B815} (2009)
  256--287, [\href{http://xxx.lanl.gov/abs/0811.2416}{{\tt 0811.2416}}].

\bibitem{Freed:1999vc}
D.~S. Freed and E.~Witten, {\it {Anomalies in string theory with D-branes}},
  {\em Asian J. Math.} {\bf 3} (1999) 819,
  [\href{http://xxx.lanl.gov/abs/hep-th/9907189}{{\tt hep-th/9907189}}].

\bibitem{Maldacena:2001xj}
J.~M. Maldacena, G.~W. Moore, and N.~Seiberg, {\it {D-brane instantons and K
  theory charges}},  {\em JHEP} {\bf 11} (2001) 062,
  [\href{http://xxx.lanl.gov/abs/hep-th/0108100}{{\tt hep-th/0108100}}].

\bibitem{Grimm:2004uq}
T.~W. Grimm and J.~Louis, {\it {The Effective action of N = 1 Calabi-Yau
  orientifolds}},  {\em Nucl. Phys.} {\bf B699} (2004) 387--426,
  [\href{http://xxx.lanl.gov/abs/hep-th/0403067}{{\tt hep-th/0403067}}].

\bibitem{Jockers:2004yj}
H.~Jockers and J.~Louis, {\it {The Effective action of D7-branes in N = 1
  Calabi-Yau orientifolds}},  {\em Nucl. Phys.} {\bf B705} (2005) 167--211,
  [\href{http://xxx.lanl.gov/abs/hep-th/0409098}{{\tt hep-th/0409098}}].

\bibitem{Jockers:2005zy}
H.~Jockers and J.~Louis, {\it {D-terms and F-terms from D7-brane fluxes}},
  {\em Nucl. Phys.} {\bf B718} (2005) 203--246,
  [\href{http://xxx.lanl.gov/abs/hep-th/0502059}{{\tt hep-th/0502059}}].

\bibitem{Berg:2004ek}
M.~Berg, M.~Haack, and B.~Kors, {\it {Loop corrections to volume moduli and
  inflation in string theory}},  {\em Phys. Rev.} {\bf D71} (2005) 026005,
  [\href{http://xxx.lanl.gov/abs/hep-th/0404087}{{\tt hep-th/0404087}}].

\bibitem{Berg:2005ja}
M.~Berg, M.~Haack, and B.~Kors, {\it {String loop corrections to Kahler
  potentials in orientifolds}},  {\em JHEP} {\bf 11} (2005) 030,
  [\href{http://xxx.lanl.gov/abs/hep-th/0508043}{{\tt hep-th/0508043}}].

\bibitem{Haack:2008yb}
M.~Haack, R.~Kallosh, A.~Krause, A.~D. Linde, D.~Lust, and M.~Zagermann, {\it
  {Update of D3/D7-Brane Inflation on K3 x T**2/Z(2)}},  {\em Nucl. Phys.} {\bf
  B806} (2009) 103--177, [\href{http://xxx.lanl.gov/abs/0804.3961}{{\tt
  0804.3961}}].

\bibitem{Berg:2011ij}
M.~Berg, M.~Haack, and J.~U. Kang, {\it {One-Loop Kahler Metric of D-Branes at
  Angles}},  {\em JHEP} {\bf 11} (2012) 091,
  [\href{http://xxx.lanl.gov/abs/1112.5156}{{\tt 1112.5156}}].

\bibitem{Berg:2014ama}
M.~Berg, M.~Haack, J.~U. Kang, and S.~Sjšrs, {\it {Towards the one-loop KŠhler
  metric of Calabi-Yau orientifolds}},  {\em JHEP} {\bf 12} (2014) 077,
  [\href{http://xxx.lanl.gov/abs/1407.0027}{{\tt 1407.0027}}].

\bibitem{Shiu:2008ry}
G.~Shiu, G.~Torroba, B.~Underwood, and M.~R. Douglas, {\it {Dynamics of Warped
  Flux Compactifications}},  {\em JHEP} {\bf 06} (2008) 024,
  [\href{http://xxx.lanl.gov/abs/0803.3068}{{\tt 0803.3068}}].

\bibitem{Martucci:2014ska}
L.~Martucci, {\it {Warping the K\"ahler potential of F-theory/IIB flux
  compactifications}},  {\em JHEP} {\bf 03} (2015) 067,
  [\href{http://xxx.lanl.gov/abs/1411.2623}{{\tt 1411.2623}}].

\bibitem{Martucci:2016pzt}
L.~Martucci, {\it {Warped K\"ahler potentials and fluxes}},  {\em JHEP} {\bf
  01} (2017) 056, [\href{http://xxx.lanl.gov/abs/1610.02403}{{\tt
  1610.02403}}].

\bibitem{Gukov:1999ya}
S.~Gukov, C.~Vafa, and E.~Witten, {\it {CFT's from Calabi-Yau four folds}},
  {\em Nucl. Phys.} {\bf B584} (2000) 69--108,
  [\href{http://xxx.lanl.gov/abs/hep-th/9906070}{{\tt hep-th/9906070}}].
  [Erratum: Nucl. Phys.B608,477(2001)].

\bibitem{Martucci:2006ij}
L.~Martucci, {\it {D-branes on general N=1 backgrounds: Superpotentials and
  D-terms}},  {\em JHEP} {\bf 06} (2006) 033,
  [\href{http://xxx.lanl.gov/abs/hep-th/0602129}{{\tt hep-th/0602129}}].

\bibitem{Grimm:2010ks}
T.~W. Grimm, {\it {The N=1 effective action of F-theory compactifications}},
  {\em Nucl. Phys.} {\bf B845} (2011) 48--92,
  [\href{http://xxx.lanl.gov/abs/1008.4133}{{\tt 1008.4133}}].

\bibitem{Frey:2002hf}
A.~R. Frey and J.~Polchinski, {\it {N=3 warped compactifications}},  {\em Phys.
  Rev.} {\bf D65} (2002) 126009,
  [\href{http://xxx.lanl.gov/abs/hep-th/0201029}{{\tt hep-th/0201029}}].

\bibitem{Witten:1996md}
E.~Witten, {\it {On flux quantization in M theory and the effective action}},
  {\em J. Geom. Phys.} {\bf 22} (1997) 1--13,
  [\href{http://xxx.lanl.gov/abs/hep-th/9609122}{{\tt hep-th/9609122}}].

\bibitem{Candelas:1990rm}
P.~Candelas, X.~C. De~La~Ossa, P.~S. Green, and L.~Parkes, {\it {A Pair of
  Calabi-Yau manifolds as an exactly soluble superconformal theory}},  {\em
  Nucl. Phys.} {\bf B359} (1991) 21--74.

\bibitem{Klemm:1992tx}
A.~Klemm and S.~Theisen, {\it {Considerations of one modulus Calabi-Yau
  compactifications: Picard-Fuchs equations, Kahler potentials and mirror
  maps}},  {\em Nucl. Phys.} {\bf B389} (1993) 153--180,
  [\href{http://xxx.lanl.gov/abs/hep-th/9205041}{{\tt hep-th/9205041}}].

\bibitem{Conlon:2016aea}
J.~P. Conlon and S.~Krippendorf, {\it {Axion decay constants away from the
  lamppost}},  {\em JHEP} {\bf 04} (2016) 085,
  [\href{http://xxx.lanl.gov/abs/1601.00647}{{\tt 1601.00647}}].

\bibitem{Greene:2000ci}
B.~R. Greene and C.~I. Lazaroiu, {\it {Collapsing D-branes in Calabi-Yau moduli
  space. 1.}},  {\em Nucl. Phys.} {\bf B604} (2001) 181--255,
  [\href{http://xxx.lanl.gov/abs/hep-th/0001025}{{\tt hep-th/0001025}}].

\bibitem{Eguchi:2005eh}
T.~Eguchi and Y.~Tachikawa, {\it {Distribution of flux vacua around singular
  points in Calabi-Yau moduli space}},  {\em JHEP} {\bf 01} (2006) 100,
  [\href{http://xxx.lanl.gov/abs/hep-th/0510061}{{\tt hep-th/0510061}}].

\bibitem{Donagi:2012ts}
R.~Donagi, S.~Katz, and M.~Wijnholt, {\it {Weak Coupling, Degeneration and Log
  Calabi-Yau Spaces}},  \href{http://xxx.lanl.gov/abs/1212.0553}{{\tt
  1212.0553}}.

\bibitem{Anderson:2013rka}
L.~B. Anderson, J.~J. Heckman, and S.~Katz, {\it {T-Branes and Geometry}},
  {\em JHEP} {\bf 05} (2014) 080,
  [\href{http://xxx.lanl.gov/abs/1310.1931}{{\tt 1310.1931}}].

\bibitem{Bielleman:2016grv}
S.~Bielleman, L.~E. Ibanez, F.~G. Pedro, I.~Valenzuela, and C.~Wieck, {\it {The
  DBI Action, Higher-derivative Supergravity, and Flattening Inflaton
  Potentials}},  {\em JHEP} {\bf 05} (2016) 095,
  [\href{http://xxx.lanl.gov/abs/1602.00699}{{\tt 1602.00699}}].

\bibitem{Ruehle:2017one}
F.~Ruehle and C.~Wieck, {\it {One-loop Pfaffians and large-field inflation in
  string theory}},  \href{http://xxx.lanl.gov/abs/1702.00420}{{\tt
  1702.00420}}.

\bibitem{Kachru:2003aw}
S.~Kachru, R.~Kallosh, A.~D. Linde, and S.~P. Trivedi, {\it {De Sitter vacua in
  string theory}},  {\em Phys. Rev.} {\bf D68} (2003) 046005,
  [\href{http://xxx.lanl.gov/abs/hep-th/0301240}{{\tt hep-th/0301240}}].

\bibitem{Balasubramanian:2005zx}
V.~Balasubramanian, P.~Berglund, J.~P. Conlon, and F.~Quevedo, {\it
  {Systematics of moduli stabilisation in Calabi-Yau flux compactifications}},
  {\em JHEP} {\bf 03} (2005) 007,
  [\href{http://xxx.lanl.gov/abs/hep-th/0502058}{{\tt hep-th/0502058}}].

\bibitem{Baume:2016psm}
F.~Baume and E.~Palti, {\it {Backreacted Axion Field Ranges in String Theory}},
   {\em JHEP} {\bf 08} (2016) 043,
  [\href{http://xxx.lanl.gov/abs/1602.06517}{{\tt 1602.06517}}].

\bibitem{Klaewer:2016kiy}
D.~Klaewer and E.~Palti, {\it {Super-Planckian Spatial Field Variations and
  Quantum Gravity}},  {\em JHEP} {\bf 01} (2017) 088,
  [\href{http://xxx.lanl.gov/abs/1610.00010}{{\tt 1610.00010}}].

\bibitem{Valenzuela:2016yny}
I.~Valenzuela, {\it {Backreaction Issues in Axion Monodromy and Minkowski
  4-forms}},  \href{http://xxx.lanl.gov/abs/1611.00394}{{\tt 1611.00394}}.

\bibitem{Blumenhagen:2017cxt}
R.~Blumenhagen, I.~Valenzuela, and F.~Wolf, {\it {The Swampland Conjecture and
  F-term Axion Monodromy Inflation}},
  \href{http://xxx.lanl.gov/abs/1703.05776}{{\tt 1703.05776}}.

\bibitem{LopesCardoso:1994is}
G.~Lopes~Cardoso, D.~Lust, and T.~Mohaupt, {\it {Moduli spaces and target space
  duality symmetries in (0,2) Z(N) orbifold theories with continuous Wilson
  lines}},  {\em Nucl. Phys.} {\bf B432} (1994) 68--108,
  [\href{http://xxx.lanl.gov/abs/hep-th/9405002}{{\tt hep-th/9405002}}].

\bibitem{Antoniadis:1994hg}
I.~Antoniadis, E.~Gava, K.~S. Narain, and T.~R. Taylor, {\it {Effective mu term
  in superstring theory}},  {\em Nucl. Phys.} {\bf B432} (1994) 187--204,
  [\href{http://xxx.lanl.gov/abs/hep-th/9405024}{{\tt hep-th/9405024}}].

\bibitem{Brignole:1995fb}
A.~Brignole, L.~E. Ibanez, C.~Munoz, and C.~Scheich, {\it {Some issues in soft
  SUSY breaking terms from dilaton / moduli sectors}},  {\em Z. Phys.} {\bf
  C74} (1997) 157--170, [\href{http://xxx.lanl.gov/abs/hep-ph/9508258}{{\tt
  hep-ph/9508258}}].

\bibitem{Brignole:1996xb}
A.~Brignole, L.~E. Ibanez, and C.~Munoz, {\it {Orbifold induced mu term and
  electroweak symmetry breaking}},  {\em Phys. Lett.} {\bf B387} (1996)
  769--774, [\href{http://xxx.lanl.gov/abs/hep-ph/9607405}{{\tt
  hep-ph/9607405}}].

\end{thebibliography}\endgroup

\end{document}